\documentclass[article]{JHEP}
\usepackage{amsmath2000,epsfig,amssymb}  % use this line to submit to Los Alamos
\title{Strings in homogeneous gravitational waves\\
and null holography}
\preprint{\hepth{0204004}\\LPTENS-02-22\\LPTHE-02-18}
\author{E. Kiritsis\\Laboratoire de Physique Th\'eorique
      de l'Ecole Normale Sup\'erieure\\
      24 rue Lhomond,
      Paris, CEDEX 05, F-75231, FRANCE\\{\rm and}\\
Department of Physics, University of Crete, and FO.R.T.H.\\
71003 Heraklion, GREECE\\
{\tt E-mail: kiritsis@physics.uoc.gr}}
\author{B. Pioline\\
LPTHE, Universit\'es Paris VI et VII, 4 pl Jussieu, \\
75252 Paris cedex 05, FRANCE\\{\tt E-mail:
pioline@lpthe.jussieu.fr}}
\abstract{Homogeneous gravitational wave backgrounds arise as infinite momentum
limits of many geometries with a well-understood holographic description.
General global aspects of these geometries are discussed. Using exact CFT
techniques, strings in pp-wave backgrounds supported by a Neveu-Schwarz flux
are quantized. As in Euclidean $AdS_3$, spectral flow and associated long
strings are shown to be crucial in obtaining a complete spectrum. Holography is
investigated using conformally flat coordinates analogous
to those of the Poincar\'e patch in AdS. It is argued that the holographic
direction is the light-cone coordinate $u$, and that the
holographic degrees of freedom live on a codimension-one screen at fixed
$u$. The usual conformal symmetry on the boundary is replaced by a
representation of a Heisenberg-type algebra $H_D\times H_D$, hinting
at a new class of field theories realizing this symmetry.  A sample
holographic computation of 2 and 3-point functions is provided
and Ward identities are derived. A complementary screen at fixed $v$ is argued
to be necessary in order to encode
the vacuum structure.}
\newcommand{\xp}{{x^+}}
\newcommand{\xm}{{x^-}}
\newcommand{\xpo}{x_0^+}
\newcommand{\xmo}{x_0^-}
\newcommand{\p}{\partial}

\newcommand{\nn}{\nonumber}

\newcommand{\Real}{\mathbb{R}}
\newcommand{\Zint}{\mathbb{Z}}
\newcommand{\Nint}{\mathbb{N}}
\newcommand{\Tr}{\mbox{Tr}}
\newcommand{\sgn}{\mbox{sgn}}

\def\bea{\begin{eqnarray}}
\def\eea{\end{eqnarray}}
\def\be{\begin{equation}}
\def\ee{\end{equation}}
\def\d{\partial}

\def\sp{\;\;\;,\;\;\;}
\def\e{\epsilon}
\def\l{\lambda}
\begin{document}
\maketitle
\section{Introduction}
Despite the success of the AdS/CFT correspondence and the recent
progress in understanding black hole entropy and holographic
bounds, the concept of holography still lacks a general background
independent formulation (see e.g. \cite{holoreview} for
recent reviews). In the standard anti-de Sitter case, progress is
hindered by the difficulty in quantizing strings in Ramond
backgrounds. For de Sitter the existence of stable string
backgrounds is itself a problem, due to the lack of
supersymmetry. Flat Minkowski space has neither of these problems,
yet its null conformal boundary makes it hard to formulate
holography.
Recently, this problem has taken a new turn with the realization that a
particular scaling limit \cite{Penrose}
of the $AdS_5\times S^5$ background, namely zooming
on a null trajectory spinning around the sphere \cite{Blau:2002dy}, leads to a
maximally supersymmetric homogeneous wave background in $D=10$
dimensions, with metric
\begin{equation}
\label{dspp} ds^2=2dudv + \sum_i dx_i^2 -\frac14 \mu^2 \left(\sum_i
x_i^2\right) ~du^2
\end{equation} where the sum runs from
$i=1$ to $8$, together with a null Ramond flux
$H=\mu~ du\wedge (dx^{1234}+dx^{5678})$ \cite{Blau:2001ne}.
Homogeneous wave backgrounds in other dimensions $D$ or supported by different
type of fluxes arise from other near-horizon geometries as discussed below.
This supersymmetric background is
amenable to exact quantization despite the existence of Ramond fluxes
\cite{Metsaev:2001bj}. The corresponding limit from the dual gauge theory
perspective was subsequently found, and a subset of the bulk string theory
states was identified with non-chiral gauge invariant operators in a very
pictorial way \cite{Berenstein:2002jq}. Most strikingly, the bulk dispersion
relation was matched to the anomalous dimension in the gauge theory side by
resumming an infinite number of non-planar diagrams.

While this scaling limit in effect gives a holographic description of these
maximally supersymmetric plane wave backgrounds, it is legitimate to wonder if
the large $J$ - fixed $(\Delta-J)$ $SU(N)$ gluons are the most appropriate
holographic degrees of freedom for this purpose. In particular, the only
manifest symmetries in this description are the $SO(4)\times SO(4)$ (for $p=5$)
unbroken symmetries, whereas the background exhibits a much larger symmetry
algebra, namely the Wigner contraction of the original $SO(2,4)\times SO(6)$
symmetry (plus fermionic generators) \cite{Hatsuda:2002xp}. Furthermore, the
scaling limit provides additional discrete symmetries (such as the exchange of
the 4 transverse directions in the $AdS_5$ part with the ones in $S^5$) which
are completely obscure in this picture. Finally, given the stringy character of
the relevant gauge theory excitations, it is rather tempting to fantasize the
existence of an exotic string theory that would be holographically dual to the
Green-Schwarz string in the bulk. Another hint that this may be so
comes from the fact that the Penrose limit for a particle in the near-horizon
geometry of a stack of D3-branes amounts to an infinite boost of the D3-branes
themselves, T-dual to the near-critical electric field limit that gives rise to
non-commutative open strings \cite{Gopakumar:2000na}.

Rather than speculating further along these lines, it may be more
enlightening to consider the pp-wave background in its own sake, and let the
geometry decide what the holographic description is, if any. This approach is
all the more useful as very similar geometries (but with different fluxes) arise
in many different systems than the near horizon limit of D3-branes. Firstly, as
already pointed out in \cite{Berenstein:2002jq}, the near horizon limit of
either M2-branes or M5-branes yields the maximally
supersymmetric wave in 11 dimensions \cite{Kowalski-Glikman:wv}. M-theory seems
to be better behaved in this exact background, in particular the flat
directions of the Matrix theory potential are lifted \cite{Berenstein:2002jq}.
It would be interesting to make progress in quantizing membranes in these
backgrounds with null curvature, but this lies outside the scope of this work
(see \cite{Bachas:2000dx} for a causally disconnected construction of
instantons interpolating between different vacua in this model).

Secondly, the maximally supersymmetric $D=10$ type IIB wave background also
arises  in the Penrose limit of systems with less supersymmetry, such as $AdS_5
\times T_{1,1}$, pointing to the existence of a subsector with $N=4$
supersymmetry in the dual $N=1$ gauge theory \cite{Itzhaki:2002kh}. Next, as
already pointed out in \cite{Berenstein:2002jq}, the $D=6$ case arises in the
Penrose limit of $AdS_3 \times S^3$, and can be supported by either Ramond or
Neveu-Schwarz flux $H=\mu~ du \wedge(dx^{12}+dx^{34})$, or by a combination of
the two. The $D=4$ case, which will be the main focus of this work, arises from
the Penrose limit of the flat NS5-brane near horizon geometry $\Real^{1,5}
\times \Real_\phi \times S^3$ \cite{chs}, by choosing a geodesic spinning along
an equator of the
sphere\footnote{This was independently noted in \cite{Gomis:2002km}.}
(it is worth noting that a geodesic along the
radial direction supporting the linear dilaton yields a purely dilatonic wave,
$\phi=\phi(u)$ with flat geometry, arguably the simplest of all plane wave
backgrounds in string theory). It is supported by a Neveu-Schwarz flux $H=\mu
du \wedge dx^{1}\wedge dx^{2}$. The same $D=4$ geometry is obtained in the
Penrose limit of the near horizon geometry of the NS5-brane wrapped on $S^2$
\cite{Gomis:2002km}, giving another example of supersymmetry enhancement in the
Penrose limit. Limits of many other backgrounds have been considered recently
\cite{penrosel}.

String theory propagation in similar gravitational wave backgrounds has been
studied previously
\cite{Nappi:1993ie,Kehagias:1994iy,Kiritsis:jk,Kiritsis:1994ij,grwav,nunez}. In
fact, the $D=4$ case is none other than the Nappi-Witten pp-wave background,
i.e. the Wess-Zumino-Witten model on the centrally extended Euclidean group in
two dimensions $E_2^c$ \cite{Nappi:1993ie}. Upon compactifying $u+v$, this
model can be viewed as an exact background with constant magnetic field
originating from the closed string sector, including the gravitational
back-reaction \cite{Russo:1994cv}. Even with non-compact $u+v$,
it shares many similarities with the Landau problem of charged particles
in constant magnetic field, with the electric charge being identified with the
light-cone momentum.

{}From the algebraic point 
of view, the solvable group $E_2^c$ can also be viewed
as the Heisenberg algebra $[P,Q]=K$ extended by the generator $J$ rotating
position and momentum, $[J,P]=Q, [J,Q]=-P$; we will denote this algebra by
$H_4$ in the sequel. The WZW model on this non-semi-simple group can also be
obtained as a limit of the WZW model on $U(1)_k \times SU(2)_{k'}$ 
where $k$ and
$k'$ are scaled to $+\infty$ and $-\infty$, respectively, with a fixed ratio
\cite{Olive:1993hk}. In particular, its central charge is that of flat space,
in agreement with the fact that there are no non-vanishing curvature invariants
in this geometry.
It was shown that the string theory on this background is exactly solvable both
in a covariant gauge \cite{Kiritsis:jk} and the light-cone gauge
\cite{Russo:1994cv, Forgacs:1995tx,Russo:2002rq} and that it preserves 16
supercharges \cite{Kiritsis:1994ij}, just as the original NS5-brane system. 

The Nappi-Witten background admits an obvious generalization in arbitrary
dimension, by considering the WZW model on the extended Heisenberg group $H_n$
\cite{Kehagias:1994iy}. This yields the pp-wave background above for $D=2n+2$,
supported by a null Neveu-Schwarz flux $H=\mu ~du \wedge \omega$ where $\omega$
is a symplectic form on the transverse space $\Real^{D-2}$. One therefore has
an exact conformal field theory description for all maximally symmetric
Cahen-Wallach spaces (see \cite{Blau:2002dy} for a review of maximally
symmetric Lorentzian spaces). As for all WZW models, these string backgrounds
are exactly solvable, and they can be mapped to free fields as spelled out in
\cite{Kiritsis:jk,Kiritsis:1994ij}.

Motivated by their ubiquitous appearance in Penrose limits of near-horizon
geometry, we will discuss general aspects of string theory and holography on
homogeneous wave backgrounds supported by a null Neveu-Schwarz flux, setting
$D=4$ in order to avoid cluttering the notation. We expect that as far as
holography goes, these models
share universal features with the ones supported by Ramond flux, of more direct
interest in the AdS context (however we will point out specific differences in
due time). In addition, these models are interesting laboratories to study
string propagation in time-dependent or cosmological backgrounds.

Let us outline the structure of this paper and summarize our main
conclusions. In section 2, we discuss general
features of the wave background \eqref{dspp}, and derive the dispersion
relation in the tree-level approximation.  In section 3, we discuss the
quantization of the RNS string in homogeneous wave backgrounds
supported by Neveu-Schwarz flux, putting
the results of \cite{Kiritsis:jk,Kiritsis:1994ij} in perspective.
In particular we give a physical discussion of the ``long string''
states required by the spectral flow introduced in \cite{Kiritsis:jk}.
Moreover, discrete symmetries analogous to the $Z_2$ symmetry
$AdS_5\leftrightarrow
S^5$ arise here as well and we show that they are symmetries
 of the quantum theory. Section 2 and 3 can be read independantly
of section 4.

In section 4, we set up a bulk-to-boundary holographic
correspondence for homogeneous wave backgrounds, irrespective of
the dimension and flux. We argue that holography is most naturally
implemented in the Poincar\'e patch \eqref{dscf}, i.e. the maximal
patch of the global geometry which is conformal to flat Minkowski
space. Under this assumption, we find that the holographic
direction is the light-cone coordinate $u$ (or equivalently
$\xp$), and that the holographic dual lives on a $D-1$ dimensional
holographic screen\footnote{This diverges from the work in
\cite{Das:2002cw} that appeared as the present work was prepared
for publication, and that advocates an Euclidean $(D-2)/2$-dimensional
dual space.} at fixed light-cone time $u$. Its induced metric is
therefore degenerate along the $v$ direction. This is in contrast
to the AdS  \cite{Witten:2001kn} and dS \cite{Strominger:2001pn}
cases, where the holographic dual lives on a time-like (resp.
space-like) Cauchy surface, and rather analogous to  Minkowski
space \cite{minkowski}. As in de Sitter, the boundary off-shell
correlators are identified with on-shell transition amplitudes
between two screens at $\xp=\pm\infty$. Thus, we expect the study
of holography in pp-wave backgrounds to shed light both on the
problems of holography in Minkowski space-time, and in de Sitter.
We determine the realization of the bulk isometries on the
boundary, and propose that the holographic description is given by
a new type of (possibly non-local) field theory with symmetry
group $H_{D}^L\times H_{D}^R$ acting as displayed in
\eqref{bounsym}, where $H_D$ is the Heisenberg algebra which
defines the pp-wave background. In the appendices, we work out the
Ward identities of two and three point functions in \eqref{f2pt}
and \eqref{f3pt}, give a sample computation for a self-interacting
scalar field in the bulk, and express the result also in the
harmonic oscillator basis.

A distinctive feature of this null holographic set-up is that, in
contrast to the AdS and dS cases, the bulk equations of motion are
first order in derivative of the holographic coordinate. While in
AdS non-normalizable (resp. normalizable) solutions are in
correspondence with boundary operators (resp. vacua), here only
the analogue of the non-normalizable modes subsist. The vacuum
structure therefore seems to be lost. As usual in light cone
quantization the details of the vacuum  are hidden. In order to
have a well-defined boundary problem it is necessary to introduce
an additional null holographic screen at fixed (infinite) $v$. The
boundary values on this screen characterize the vacuum structure
of the theory on the holographic screen at fixed $u$. Only trivial
boundary conditions at $v\to-\infty$ leave the bulk symmetry
unbroken.

\section{Geometry of homogeneous wave backgrounds}

\subsection{The homogeneous wave as a group manifold}

As we explained in the introduction,
the homogeneous plane wave metric \eqref{dspp},
supported by a Neveu-Schwarz flux, can be constructed as the manifold of the
$D-1$ dimensional Heisenberg group, extended by a generator rotating the
positions and momenta:
\begin{equation}
H_D:\quad [P_i,P_j]=\omega_{ij} K\ ,\quad [J,P_i]=\omega_{ij} P_k
\end{equation}
where $K$ is a central element. This can also be thought as the
non-commutative phase space for a particle in $D-2$ dimensions
in a constant magnetic field $B=\omega_{ij} dx^{ij}$, with
$J$ generating a rotation in each of the proper planes of $B$.
The center of the group is generated by $K$ and the Casimir
$C=P_T^2+2 JK$ where $P_T^2=\sum P_i^2$.
Restricting for simplicity to $D=4$, and letting
$\omega_{12}=1$, the group has the (non-unitary) matrix representation
\be
\label{gmat}
g(u,v,a_1,a_2)=e^{a_1P_1+a_2P_2}e^{uJ+vK}=
\begin{pmatrix}
1& a_1 \sin u-a_2 \cos u  & a_1\cos u + a_2\sin u& 2v\\
0& \cos u& -\sin u& a_1\\
0& \sin u& \cos u& a_2\\
0& 0& 0& 1
\end{pmatrix}
\ee
In this presentation the generator $J$ appears to be compact,
while the other ones are non-compact. We will consider mostly the
universal cover in order to avoid closed light-like curves. However
for some DLCQ applications it may be desirable to keep $u$ compact. We
will argue later that, for compact $u$, string theory is only
consistent if $u$ has $2\pi/n$ periodicity  (after setting the mass scale
of the background to 1).

A bi-invariant metric can be obtained as $ds^2=\langle g^{-1}dg,
g^{-1}dg \rangle$, where $\langle \cdot,\cdot \rangle$ is an
invariant symmetric form on the Lie algebra. The trace
in the above matrix representation yields a degenerate metric
$\Tr( g^{-1}dg g^{-1}dg)=-2du^2$, however there is another
invariant form, $\langle P_i,P_j \rangle=\delta_{ij},
\langle J,K \rangle=1$
with other components vanishing. This yields a non-degenerate metric,
\be
\label{dsppa}
ds^2=2du dv +da_1^2+da_2^2+ (a_2da_1-a_1 da_2) du\ .
\ee
This is supported by the Neveu-Schwarz flux $H=du\wedge da_{12}$. In this form,
one recognizes a constant magnetic background $F=da_{12}$ for the Kaluza-Klein
gauge field $g_{\mu u}$ -- except for the fact that the $u$ coordinate is null.
The Laplacian in these ``magnetic'' coordinates reads \be \label{dsmag}
\Delta=2\p_u\p_v+\left(\p_1-\frac{1}{2}a_2\p_v\right)^2+
\left(\p_2+\frac{1}{2}a_1\p_v\right)^2 \ee so that the free wave equation
$(\Delta-m^2)\phi=0$ reduces to the Schrodinger equation for a particle of
charge $p_+$ (the momentum along $v$) in a constant magnetic field. Changing to
the rotating frame at the Larmor frequency, $a_1+ia_2=e^{iu/2}(x_1+ix_2)$,
leads\footnote{Changing to the frame rotating at twice the Larmor frequency,
$a_1+ia_2=e^{iu}(b_1+i b_2)$, leads back to a magnetic background but with
opposite magnetic field. It amounts to passing $e^{a_i P_i}$ through
$e^{uJ+vK}$ in \eqref{gmat}.} to the more familiar metric \be \label{dspp2}
ds^2=2du dv + dx_1^2 +dx_2^2 -\frac14 (x_1^2+x_2^2) ~du^2 + ds^2_{\perp}\ ,
\ee
supported by the Neveu-Schwarz flux $H=du\wedge dx^{12}$.
The equations of motion are easily seen to be satisfied, since the only
non-vanishing component of the Ricci tensor, $R_{uu}=1/2$, is compensated by
the NS flux $H_{uij}H_u^{\;ij}$.

\subsection{Isometries and geodesics}
The metric \eqref{dsmag} or \eqref{dspp2} is
by construction invariant under $H_{4L}\times H_{4R}$ acting from the left and
the right on the group $H_4$ itself, respectively. The group law is easily
computed, e.g. using the Baker-Campbell-Hausdorff formula
\cite{Figueroa-O'Farrill:1999ie}, or using the matrix representation, \be
g(u_1,v_1,z_1)g(u_2,v_2,z_2)=g\left(u_1+u_2, v_1+v_2-\frac12 \Im(z_1
e^{-i(u_1+u_2)/2}\bar z_2), z_1 e^{-iu_2/2}+z_2 e^{iu_1/2}\right) \ee in
particular the inverse is $g(u,v,z)^{-1}=g(-u,-v,-z)$. Since the element $K$ is
central, the resulting group of isometries is seven-dimensional.
The Killing vectors associated to the left and right action follow \footnote{In
our conventions, the right generators have opposite commutation relations to
the left ones.} from the group law, and read:
\begin{eqnarray}
K&=&\p_v\\ \nn
J_L&=&\p_u + (x_1 \p_2-x_2 \p_1)/2 \\
J_R&=&\p_u - (x_1 \p_2-x_2 \p_1)/2 \nn
\end{eqnarray}
\begin{eqnarray*}
P_{1L}&=&\frac12\left(x_2\cos\frac{u}{2} + x_1\sin\frac{u}{2}\right)\p_v
+ \cos\frac{u}{2} \p_1  -\sin\frac{u}{2} \p_2\\
P_{1R}&=&\frac12\left(-x_2\cos\frac{u}{2}+ x_1\sin\frac{u}{2}\right)\p_v
+ \cos\frac{u}{2} \p_1+ \sin\frac{u}{2} \p_2\\
P_{2L}&=&\frac12\left(-x_1\cos\frac{u}{2}+ x_2\sin\frac{u}{2}\right)\p_v
+ \sin\frac{u}{2} \p_1+ \cos\frac{u}{2} \p_2\\
P_{2R}&=&\frac12\left(x_1\cos\frac{u}{2} + x_2\sin\frac{u}{2}\right)\p_v
-\sin\frac{u}{2}\p_1+ \cos\frac{u}{2} \p_2
\end{eqnarray*}
where $\p_i:=\p/\p x_i$. In particular, $K$, $J_L$ and $J_R$
are null Killing vectors. The combination $J_L+J_R=2\p_u$ is a timelike
Killing vector, and will play a special role in our discussion
as the holographic generator.

\subsection{Geodesics and waves}
It is of interest to investigate the free trajectories in this background. This
is especially simple in this homogeneous case, since the geodesic flow is
generated by the left (or right) action of a one-parameter subgroup of $H_4$,
$g(\tau)=\exp(2\tau(p_+ J+p_- K +p_1 P_1+p_2 P_2))$ depending on the initial
momentum. Starting without loss of generality from the origin $(u,v,z)=(0,0,0)$
of the group, one finds that the geodesics are parameterized by \bea
u(\tau)&=&2p_+ \tau \nn\\
v(\tau)&=&\left( 2p_- +\frac{p_1^{2} + p_2^{2}}{p_+} \right)\tau
-\frac{\sin 2p_+  \tau }{2p_+^2}(p_1^{2} + p_2^{2}) \label{geo}\\
z(\tau)&=&2\sin( p_+ \tau)~(p_1 + i p_2) / p_+\nn \eea
The particle therefore
oscillates in the harmonic potential well in the transverse $z$ direction with
a period $\Delta \tau=2\pi/p_+$ (the period is $\pi/p_+$ in $w$ coordinates).
During one period, the particle moves linearly by $\Delta u=4\pi$ in the $u$
direction, and by $\Delta v =4\pi p^2/2p_+^2$ in the $v$ direction, where
$p^2=2p_+ p_-+p_1^2+p_2^2$ is the invariant momentum square at $\tau=0$. The
mean velocity $\Delta v/\Delta u=p^2/2p_+^2$ therefore depends on the
type of trajectory under consideration: massive particles have a net forward
motion along the $v$ axis (as $u$ increases) while space-like trajectories go
backward in $v$. Massless particles on the other hand are confined in the $v$
direction on an interval of length $(p_1^2+p_2^2)/p_+^2$. In either case, the
Zitterbewegung-like motion along $v$ forbids to take it as a time coordinate,
in contrast to $u$ which is a single valued function of the proper time $\tau$.
If on the other hand the light-cone momentum $p_+$ vanishes, the potential well
disappears, and the motion is linear in the $(v,x_1,x_2)$ plane with fixed $u$.
This in general corresponds to a space-like geodesic.
Null geodesics
with $p_+=0$ must also have $p_1=p_2=0$, and correspond to
massless particles traveling
parallel to the wave. Starting away from the bottom of the potential changes
the details of the motion in the transverse plane, but the discussion of the
motion in the $(u,v)$ directions still applies.

Having at hand the general form of the trajectories, one readily
obtains the geodesic (Lorentzian) distance $r$ between any two points,
\be
r^2 =2(u-u')(v-v')+ \frac{u-u'}{2\sin\left(\frac{u-u'}{2}\right)}
\left[
\left(x_1^2+x_2^2+x_1^{'2}+x_2^{'2}\right)  \cos\left(\frac{u-u'}{2}\right)
-2(x_1 x_1'+x_2 x_2')\right]
\label{distance}\ee
The geodesic distance therefore goes to infinity
when $u-u'=2\pi n$, $n\in \Zint$,
unless $(x_1,x_2)=(x_1',x_2')$ for $n$ even, or
$(x_1,x_2)=(-x_1',-x_2')$ for $n$ odd. This agrees with the periodicity
of the trajectories in the transverse plane.

These features of the classical trajectories carry over
to the propagation of waves in this background. Scalar waves
satisfy $(\Delta-m^2)\phi=0$ where the Laplace operator can be
computed either from the metric \eqref{dspp} or from the Casimir
of the left or right generators, \be\label{laphar}
\Delta=2\p_u\p_v+\frac14(x_1^2+x_2^2)\p_v^2+(\p_1^2+\p_2^2)+\Delta_\perp
\ee
 Acting on a
plane wave $e^{i (u p_- + v p_+)}$ with non-zero $p_+$, one gets
the Hamiltonian for a harmonic oscillator in two dimensions with
frequency $|p_+|/2$. Its ground state is the Gaussian wave function
\be \phi(u,v,z)=\exp\left[ i (p_+ v + p_- u + p_\perp x_\perp) -
\frac14 |p_+|z \bar z\right] \ee while arbitrary states are
obtained by acting $n_L$ times (resp. $n_R$) with the creation
operators $P_L^-$ and $P_R^+$, \bea P^+_{L}&=&e^{iu/2}(4\p_{\bar
z}-i z \p_v) \ ,\quad
P^-_{L}=e^{-iu/2}(4\p_{z}+ i\bar z \p_v)\\
P^+_{R}&=&e^{-iu/2}(4\p_{\bar z}+i z \p_v) \ ,\quad
P^-_{R}=e^{iu/2}(4\p_{z}-i\bar z \p_v) \nn \eea where $P^{\pm}=P_1\pm i P_2$,
$z=x_1+ix_2, \p_z=(\p_1-i\p_2)/2$. The dispersion relation is easily seen to be
\be \label{disp} 2 p_+ p_- +  |p_+| (n_L + n_R + 1) + p_\perp^2 + m^2 =0 \
,\quad \ee while the helicity in transverse place is $h=\sgn(p_+)(n_L-n_R)$. We
will re-derive this spectrum more systematically in the next section. For
$p_+=0$, the wave equation reduces to the Laplace operator in transverse space,
and has no solution for positive $m^2$. For $m^2=0$, the solution is an
arbitrary profile in $u$, independent of the remaining coordinates.
It is interesting to note from the dispersion relation \eqref{disp} that
massless states can correspond to either $p_+=0$ and arbitrary $p_-$, or to
$p_-=\sgn(p_+)(n_L+n_R+1/2)$ and arbitrary $p_-$ (with $p_\perp=0$ of course).

\EPSFIGURE{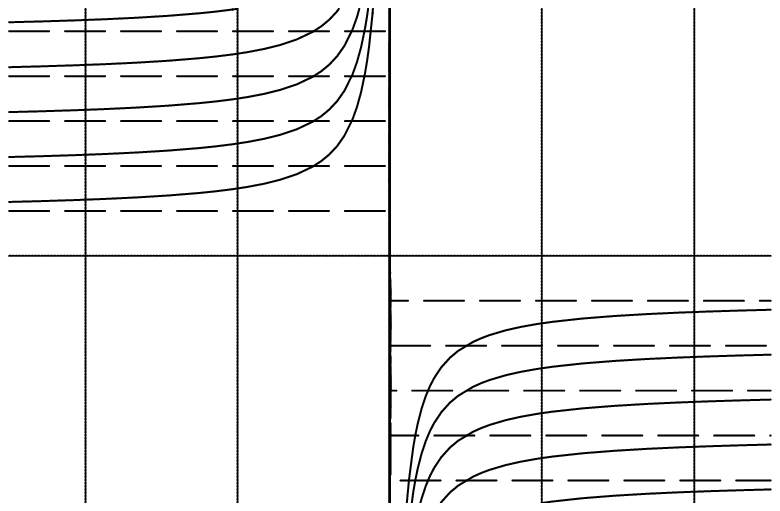,height=6cm}{Dispersion relation in the $(p_+,p_-)$
plane.\label{sugrad}}

For zero mass (in dotted lines), the spectrum of $p_-$ takes discrete positive
values (resp. negative) at $p_+<0$ (resp. $p_+>0$), and forms a lowest (resp.
highest) weight representation of $H_{4L}\times H_{4R}$. At $p_+=0$, the
spectrum of $p_-$ is continuous, with each value transforming as a singlet. For
$m^2>0$, $|p_-|$ develops a mass gap, and there are no states at $p_+=0$
anymore. Vertical lines denote integer $p_+$. For the bosonic string, the
spectrum is further tilted and there is a tachyonic state.

\subsection{Representation theory of the extended Heisenberg group $H_D$}
The spectrum found above should fall into unitary irreducible representations
of the left and right $H_4$ actions. The unitary irreps of $H_4$ were
constructed in \cite{Streater} and from a more applied point of view in
\cite{Kiritsis:jk}. We will briefly recall their construction, in notations
consistent with this paper (in particular all generators $P_{1,2},K,J$ are
taken to be anti-hermitian).

The unitary irreps can be constructed by
diagonalizing simultaneously the commuting operators $C,K,J$: \be J |j,t\rangle
= i~j  |j,t\rangle\ ,\quad K |j,t\rangle = i~t  |j,t\rangle\ ,\quad C
|j,t\rangle = C  |j,t\rangle\ \ee with $j$ and $t$ real. $K$ and $C$ are
constant in a given irreducible representation. Since $P_1^2+P_2^2$ is negative
definite, we must have $C+2jt\leq 0$. The operators $P^{\pm}=P_1\pm i P_2$ can
be thought of as raising and lowering operators for $J$, as $[J,P^\pm]=\mp i
P^\pm$. Choosing an orthonormal basis and making a choice of phases, they act
as \be P^\pm |j,t\rangle = \sqrt{-[C+t(2j\mp 1)]} ~ |j\mp 1,t\rangle \ee We
then have the following representations:

\begin{itemize}
\item[(I)] Representations without highest nor lowest weight have $j$
ranging over $j_0 + \Zint$, and $j_0$ can be chosen in $[0,1[$
without loss of generality. In order that $C+2jt\leq 0$ for all
states, we must have $t=0$ and $C<0$.

\item[(II)] Representations with lowest weight $P^-|j,t\rangle=0$
have $C=-t(2j+1)$. The negativity of $P_1^2+P_2^2$ implies $t>0$.
The spectrum of $J$ is bounded from above, $[J]=j-\Nint$. The
Casimir $C$ is positive for $j<-1/2$.

\item[(III)] Representations with highest weight $P^+|j,t\rangle=0$
have $C=-t(2j-1)$. The negativity of $P_1^2+P_2^2$ implies $t<0$.
The spectrum of $J$ is bounded from below, $[J]=j+\Nint$. The
Casimir $C$ is positive for $j>1/2$.
\item[(IV)] Representations with both highest weight $j$ and lowest weight
$j'$ must have $j=j'$: there are therefore one-dimensional only.
They have $C=t=0$, and $j$ is any real number.
\end{itemize}

The spectrum of the lowest and highest weight representations II and III can be
summarized by $C=-2tj-|t|$ and $J=j-\sgn(t)\Nint$. For a fixed value of the
charges $t$ and $J$, the Casimir is then \be \label{cas} C=-2t J-(2\Nint+1)|t|\
. \ee In particular, irrespective of the sign of $t$, it is always discrete
and, for a fixed $J$, bounded from above. As $t$ goes to zero, the Casimir goes
to zero. For $t=0$ strictly, we have a singlet at $C=0$ (type IV), and a
continuum of states with $C<0$ (type i).

 Let us now go back to the spectrum of
the Laplacian in the background \eqref{dspp}, and fit it into representations
of $H_{4L}\times H_{4R}$. The eigenvalues of $K, J_L, J_R$ are identified as
$t_L=t_R=p_+, j_L=p_- - h/2$, $j_L=p_- + h/2$, while the Casimir $C=C_L=C_R$ is
identified with the mass squared. For representations with highest or lowest
weight on both sides, \eqref{cas} implies $j_L=j_R$ on the highest (lowest
weight), which should therefore have zero helicity. If $u$ was taken to be
compact, this would further imply that $j_L+j_R=2p_-$ is quantized. The sum of
\eqref{cas} on the left and right side reproduces the dispersion formula
\eqref{disp}, while the difference identifies $h=\epsilon(p_+)(n_L-n_R)$.

The spectrum of the light-cone Hamiltonian $p_-$ as a function of the
light-cone momentum $p_+$ for a fixed value of $m^2$ (or more accurately
$m^2+p_\perp^2$) is represented in Figure 1. For $m^2=0$, we have a highest
weight representation of type II at $p_+<0$, a lowest weight representation of
type III at $p_+>0$, and an infinity of type IV singlet representations at
$p_+=0$, arbitrary $p_-$. The light-cone energy of the type II and III
representations is independent of the momentum. For $m^2>0$, we have only type
II and III representations at $p_+>0$ and $p_+<0$, respectively. In contrast to
flat space, the light-cone Hamiltonian has a mass gap, $|p_-|>1/2$.
Representations of type IV exist only for tachyonic mass, and will not occur in
the superstring case.

\section{Strings in Neveu-Schwarz homogeneous pp-waves}

String theory in the  Nappi-Witten background (or its higher dimensional
generalizations) can be solved in the light-cone gauge, where the transverse
scalars become free massive fields \cite{Russo:1994cv,Forgacs:1995tx}.
Alternatively, one may avoid the light-cone gauge and make use instead of the
$H_{DL}\times H_{DR}$ current algebra \be \label{ca}
J_{a}(z)J_{b}(z')=\frac{G_{ab}}{(z-z')^2}+
{f_{ab}}^{c}\frac{J_{c}(z')}{(z-z')}+ {\rm regular} \ee where $G_{ab}$ is the
non-degenerate invariant symmetric form on the Lie algebra introduced before,
$\langle P_1,P_2\rangle= \langle J,K\rangle=1$, rescaled by a factor of the
affine level $k$.

The analogue
of the affine-Sugawara stress tensor, satisfying the Master equation
\cite{Halpern:1989ss}, is then given by $T=-L^{ab}J_aJ_b$ where
$L^{ab}=G^{ab}/2+\tilde G^{ab}/(2k^2)$, where $\tilde G^{ab}$ is the degenerate
invariant symmetric form with only non-vanishing element $\tilde G^{KK}=1$. The
affine level $k$ can be set to one by rescaling $\alpha'$.

The left-moving and right-moving currents can be computed from $J_L(z)=\langle
g^{-1}\p g , J \rangle, J_R(z)=\langle \bar\p g g^{-1}, J \rangle$,
respectively. In the coordinate system (\ref{gmat}), we find that the
left-moving currents take a simple expression, \bea \label{lmc}
K_L=\p u\ &,&\quad J_L=\p v +\frac12 (a_2 \p a_1-a_1 \p a_2)\ ,\quad\\
P^+_L=e^{-iu} \p w\ &,&\quad P^-_L=e^{iu} \p \bar w\ ,\quad
\eea
whereas the right-moving currents are more awkward,
\bea
K_R=\bar\p u\ &,&\quad J_R=\bar\p v +\frac12 (a_2 \bar\p a_1-a_1 \bar\p a_2)
-\frac12(a_1^2+a_2^2)\bar\p u\ ,\quad\\
P^+_R=\bar\p w-iw\bar\p u\ &,&\quad P^-_R=\bar\p \bar w+i \bar w
\bar \p u \eea
Changing to the rotating frame at twice the Larmor frequency would reverse the
situation. In fact, it was shown in \cite{Kiritsis:jk} that the current algebra
has the free boson representation, reminiscent of \eqref{lmc}, \bea
\label{fb} K=\p u\ ,\quad J=\p v\ ,\quad P^+_L=e^{-iu} \p w\ ,\quad
P^-_L=e^{iu} \p \bar w\ ,\quad \eea where $P^{\pm}_{L}=(P_1\pm iP_2)/\sqrt{2}$.
The scalars $(u,v,w,\bar w)$ have free boson propagators, \be \langle
x^{\mu}(z) x^{\nu}(z')\rangle~=~ \eta^{\mu\nu}\log(z-z') \ee with metric
$\eta_{\mu\nu} dx^\mu dx^\nu=2dudv+2dw d\bar w$.

This free-field representation amounts to choosing different coordinate systems
for the left and right movers, namely $w=a_1+ia_2$ on the left and $w=b_1+ib_2$
on the right, with $e^{-iu/2}(a_1+ia_2)=x_1+ix_2=e^{iu/2}(b_1+i b_2)$. In
addition, one drops the singular helicity operator in $J(z)$. One may then
check that the operators \eqref{fb} satisfy the current algebra \eqref{ca}. The
stress tensor then takes the free boson form, $T=-{1\over
2}\eta_{\mu\nu}\partial x^{\mu}\partial x^{\nu}$.

With these notations, the vertex operator $:e^{i p_- u + i p_+v}:$ generates
the state $|p_+,p_-\rangle$ with $t=p_+$, $j=p_-$ and $L_0=-p_+p_-$. This
free-field resolution of the algebra generalizes in an obvious fashion to all
Cahen-Wallach (CW) spaces with NS-NS two-form, by adding extra pairs of raising
and lowering operators $P_{Li}^\pm$ expressed in terms of free complex
coordinates.

\subsection{Chiral primaries and vertex operators}

As shown in \cite{Kiritsis:jk} type I representations correspond to the states
$|p_+,p_-\rangle$ with $p_+=0$ and all oscillator states generated by
$P^{\pm}$. We should note that such states are invisible in the light-cone
gauge. It turns out that only tachyonic or massless states may belong to this
class. In particular, for the superstring, where tachyons are absent, the only
physical states in this class are the ground-states $|p_+=0,p_-\rangle\otimes
|0\rangle$ where $|0\rangle$ is the vacuum of the extra six-dimensional CFT.
They correspond to the massless modes traveling with the wave. Their tree
level interactions are the same as in Minkowski space.

Type II primaries can be represented as \cite{Kiritsis:jk} \be \label{rii}
R^{II}_{p_{+},p_{-}}=\exp\left[i(p_{-}u+p_{+}v)\right] H_{p_{+}}(x^{1},x^{2}).
\ee Here we have $p_{+}>0$. The highest weight condition  implies that \be \d w
(z) H_{p_{+}}(z')\sim (z-z')^{-1-p_{+}}\ ,\quad \d \bar w(z) H_{p_{+}}(z')\sim
(z-z')^{p_{+}},\label{twist}
\ee
It is apparent from the OPEs above  that the vertex operator $V^{II}_{p_{+}}$
is an operator that twists the transverse plane coordinates via the
transformation \be w(e^{2\pi i}z)=e^{-2\pi ip_{+}}w(z)\;\;,\;\;\bar w(e^{2\pi
i}z)=e^{2\pi ip_{+}}\bar w(z). \label{twist2}
\ee
Equivalently, they modify the periodicity of the free boson $w$ along the
world-sheet spatial coordinate
\cite{Kiritsis:jk,Kiritsis:1994ij,Russo:1994cv,Forgacs:1995tx}. Note that in
general the twist need not be a fractional number. The conformal weight of this
twist field is
%\footnote{Note that in our conventions the $L_0$ eigenvalues of a
%unitary CFT are negative}
${p_{+}}\left(1-{p_{+}}\right)/2$ so that the total $L_0$
conformal weight is \be \Delta= p_{+}p_{-} + {1\over
2}{p_{+}}\left(1-{p_{+}}\right)\ ,\qquad p_{+}>0 \label{cw} \ee in
agreement with the Casimir value of the classical $H_4$ algebra,
up to the stringy correction term quadratic in $p_+$. These states
correspond to the confined Landau-type harmonic oscillator states
in the transverse plane (they can of course carry momentum in the
extra $10-D$ directions). The energy formula \eqref{cw} is only
valid for $0\leq p_+<1$, since as $p_+\to p_++1$ the free boson
$w$ is twisted by the same amount $e^{-2\pi i p_+}$.

Let us expand the free boson $\d w$, $\d \bar w$ in oscillator modes
$a_{k+p_+}$ and $\bar a_{k-p_+}$. The action of the affine generators $P_n^\pm$
can then be written as \bea P^{+}_{k}~|p_+,p_-\rangle\otimes |H_{p_+}\rangle
&=&|p_+,p_--1\rangle\otimes a_{k+p_+}|H_{p_+}\rangle \\
P^{-}_{k}~|p_+,p_-\rangle\otimes |H_{p_+}\rangle
&=&|p_+,p_-+1\rangle\otimes \bar a_{k-p_+}|H_{p_+}\rangle  \nn
\eea
Due to (\ref{twist}), the oscillators $a_{k+p_+}$ for $k\geq 1$
and $\bar a_{k-p_+}$ for $k \geq 0$ annihilate the twisted vacuum,
while the other ones act as creation operators. A general state
can be written as \be \prod_{k=0}^{\infty} ( a_{-k+p_+}  )^{N_k}
\prod_{\bar k=1}^{\infty} ( \bar a_{-\bar k-p_+}  )^{\bar N_{\bar
k}} |p_+,p_- - \sum N_k + \sum \bar N_{\bar k} \rangle \otimes
|H_{p_+}\rangle \ee Its conformal dimension is
\bea L_0&=&p_+
P_- + \frac12 p_+(1-p_+) + \sum_{k=0}^{\infty} ( k-p_+  ){N_k}
+ \sum_{\bar k=1}^{\infty} ( \bar k+ p_+  ){N_{\bar k}}    \nn\\
&&=p_+ P_- +  p_+ n + \frac12 p_+(1-p_+) + N
\label{totl0}
\eea
where $N=  \sum_{k=0}^{\infty} k N_k
+ \sum_{\bar k=1}^{\infty} \bar k N_{\bar k}$ is the total
oscillation number, $n=\sum N_k - \sum \bar N_{\bar k}$
and $P_-=p_--n$ is the light-cone energy of the resulting state.
Type III primaries are conjugate to type II primaries. They are essentially
constructed like the type II but starting on the $|p_+<0,p_-\rangle$ state
\cite{Kiritsis:jk}. They have the same conformal weights as in
(\ref{cw}) but with $p_+\to |p_+|$.

\subsection{Spectral flow and long strings}

As we have seen, the primary field \eqref{rii} only gives rise to states with
$0\leq p_+<1$. Increasing $p_+$ by one amounts to performing the spectral flow
\bea J_m&\to& J_m\ ,\quad K_m\to K_m+iw\delta_{m,0}\ ,\quad P^{\pm}_{m}\to
P^{\pm}_{m\pm w} \\
L_0&\to& L_0-iw\left(K_0-J_0\right)-\frac12{w(1-w)}\delta_{m,0} \eea
When $w\in
\Zint$ we obtain an isomorphic algebra, but highest weight vectors are now
defined to be eigenmodes of different Cartan generators, $J_0$ and $K_0+iw$,
and annihilated by operators $P^\pm_{n\pm w}$ with $n\geq 0$. One therefore
obtains different representations, with conformal dimension \be
\Delta(w)=\Delta(0)-w(p_+-p_-)+\frac12 w(1-w) =p_+' p_- +\frac12 p_+' (1-p_+')
\label{cw1} \ee where $p_+'=p_++w$ is the new momentum.
The formula \eqref{cw} appears to still be valid for $p_+>0$
states outside the range $0<1<p_+$, but with the understanding
that a representation of arbitrary $p_+>0$ is really obtained from
the standard type II representation of light-cone momentum $p_+
\mod 1$ by flowing by $[p_+]$ units. A similar statement applies
to the $p_+<0$ states.

It is useful to understand the geometric meaning of these new representations,
in particular in order to determine the allowed spectral flows. Starting from a
classical solution $g(\tau,\sigma)=g(\tau+\sigma)g(\tau-\sigma)$ of the WZW
model, the spectral flow by $w_L$ units on the left and $w_R$ on the right acts
as \be g(\tau,\sigma) \to e^{w_L(\tau+\sigma) J} g(\tau,\sigma)
e^{w_R(\tau-\sigma)J} \ee In terms of the string embedding coordinates, this
amounts to transforming \bea
u&\to& u + (w_L+w_R)\tau+(w_L-w_R)\sigma \\
v&\to&v\\
x_1+ix_2 &\to& e^{\frac{i}2(w_L+w_R) \sigma+
\frac{i}2(w_L-w_R)\tau} (x_1+ix_2)
\eea
Starting from the generic particle trajectory
$g(\tau)=e^{2\tau(p_- J + p_+ K + p_i P_i)}$
that we analyzed in \eqref{geo}, one obtains string worldsheets
parameterized by
\bea
u(\tau,\sigma)&=&(2p_+ + w_L+w_R)\tau+(w_L-w_R)\sigma  \nn \\
v(\tau,\sigma)&=& \frac{2 p_- p_+ +  p_1^{2} +
p_2^{2}}{p_+}\tau
- \left[p_1^2+p_2^2\right] \frac{\sin(2p_+)}{2p_+^2} \\
z(\tau,\sigma)&=&2\sin(p_+\tau)~e^{\frac{i}2(w_L+w_R) \sigma+
\frac{i}2(w_L-w_R)\tau} \left[ p_1 + i p_2 \right] / p_+\nn \eea
Unless the coordinate $u$ is compact, one should therefore only
allow for symmetric spectral flow $w_L=w_R:=w$. The resulting
configuration is a string winding $w$ times around the bottom of
the potential in transverse space. This state carries charges
$P_-=p_-$, $P_+=p_++w$, and has conformal dimension $L_0=\bar
L_0=\frac12(p_1^2+p_2^2)+ (p_++w)p_-$, to be equated to  the
conformal weight in the flat transverse directions $h$. The
spectral flow from a time-like geodesic (resp. space-like)
therefore yields a state of $p_-<h/w$ (resp. $p_->h/w$). The
light-cone energy is extremized when $p_i=0$, where the string
degenerates to a particle sitting at the bottom of the harmonic
potential well. On the other hand, one may consider the spectral
flow from a $p_+=0$ geodesic, with momentum $(0,p_-,p_1,p_2)$ and
starting at an arbitrary position $(0,0,x_1,x_2)$ in transverse
space. One obtains after a symmetric flow \bea
u(\tau,\sigma)&=&2 w \tau \nn \\
v(\tau,\sigma)&=&2 p_- \tau\\
z(\tau,\sigma)&=&e^{iw \sigma}
\left[ (x_1 + i x_2) +  \tau (p_1+i p_2) \right] \nn
\eea
This state carries charge $J_L=J_R=w, K_L=p_-, K_R=p_- + p_1
x_2-p_2 x_1$, and has conformal dimensions
$L_0=\frac12(p_1^2+p_2^2)+p_- w, \bar L_0=L_0+w(p_1 x_2-p_2 x_1)$.
The Virasoro conditions $L_0+\bar L_0=2h, L_0-\bar L_0=0$
determine the momenta $p_i$ in terms of the initial position
$x_i$. The resulting state \bea z(\tau,\sigma)&=&e^{iw \sigma}
(x_1 + i x_2) \left[ 1 \pm 2 \tau \sqrt\frac{2(h-p_-
w)}{x_1^2+x_2^2}
 \right]
\eea
describes a closed string of winding number $w$ around the bottom
of the potential, contracting or expanding linearly with time. At
the critical value $p_-=h/w$, this becomes a static string, whose
radius can stretch in the transverse directions at no cost in
energy. Its light-cone momentum $p_+$ is fixed to be integer. This
string is the analogue of the long string appearing in Euclidean
$AdS_3$ \cite{Maldacena:2000hw}. Upon compactifying $u$, one
obtains for $w_L\neq w_R$ strings that wind around $u$ and have
correlated angular momentum in the transverse direction. Note that
the radius of $u$ is fixed to $2\pi$, since $w_L-w_R$ is required
to be integer. It would be interesting to relate these extended
states to the thermal instabilities found in \cite{Russo:1994ew}.

\subsection{String spectrum}

We can now describe the spectrum of physical states of string
theory in the background $H_4\times C_6$ where $C_6$ stands for
the extra six-dimensional background. We will assume that $C_6$ is
the Euclidean space $\Real^6$, as e.g. in the Penrose
limit of the NS5-brane case. We choose a physical
gauge \footnote{This is not exactly the
light-cone gauge since here states with $p_+=0$ are retained.} by
dropping the non-trivial oscillator modes of the currents
$J_L,K_L,J_R, K_R$.

The level zero left- and right-moving generators are identified as
$K_{L0}=K_{R0}=p_+$, $J_{0L}+ J_{0R}=2p_-$, $J_{0L}-J_{0R}=h$ where $h$ is the
(integer) helicity in the transverse plane.
The generic physical state is generated from
the basic level zero states \be |p_+,p_-,h;n_L,n_R \rangle\otimes
|p_T\rangle\ee by the action of the negative modes of the
$H_4^L\times H_4^R\times U(1)^6_L\times U(1)^6_R$ current algebra
($p_T$ labels momentum on $\Real^6$). Thus, taking into
account the spectral flow, the physical state
conditions in the bosonic string case are
\bea
L_0&=&p_+^L p_-^L+ p_+^L n_L + \frac12 p_+^L(1-p_+^L)+\frac12
p_\perp^2 +N_L=1 \\
\bar L_0
&=&p_+^R p_-^R+ p_+^R n_R + \frac12 p_+^R(1-p_+^R)+\frac12
p_\perp^2 +N_R=1 \nn
\eea
where $p_+^L=p_+ + w_L$, $p_+^R=p_+ + w_R$,
$p_-^L=p_- - h/2$, $p_-^R=p_- + h/2$, and $n_{L,R}, N_{L,R}$ are
defined as below \eqref{totl0}.

Let us first assume that neither of the light-cone directions are compact.
The spectral flow has to be symmetric, hence $p_+^L=p_+^R=p_+$.
The matching condition follows by taking the difference $L_0-\bar L_0$:
\be
\label{match}
p_+ (h-\sgn(p_+)( n_L-n_R))=N_L-N_R
\ee
Since in any consistent CFT the conformal spin must be integer, and since $p_+$
is continuous, we must impose from the start the matching condition
$h-\sgn(p_+)( n_L-n_R))=0$, identical to the tree-level answer. This further
implies $N_L=N_R$ for the left and right excitation numbers of the string.
Taking now the sum $L_0+\bar L_0$, we find the physical mass spectrum of the
closed bosonic string in the Nappi-Witten background, \be \label{disp1} 2 p_+
p_- +|p_+| (n_L + n_R) + |p_+|(1-|p_+|) + p_\perp^2 + N_L+N_R -2 =0 \ ,\quad
\ee In this expression, $p_+$ runs from $-\infty$ to $+\infty$, keeping in mind
the fact that states outside the range $-1\leq p_+\leq 1$ are
``long string'' states obtained by spectral flow from the domain
$|p_+|<1$. This differs by a quadratic term in $p_+$ from the
tree-level dispersion relation \eqref{disp}. Reinstating units,
the correction is $p_+(1- \mu \alpha' p_+)$, hence it can be
thought as a one-loop correction in the $\sigma$ model. Its effect
is to tilt the spectrum in the $(p_+,p_-)$ plane. The quadratic
correction cancels in the supersymmetric case (e.g. relevant for
the NS5-brane) due to the fact that the fermions $\psi$ and $\bar
\psi$ are twisted by the same amount as their bosonic partner $w$
so that the supercurrent remains invariant \cite{Kiritsis:1994ij}.
This structure should be contrasted with the case of Ramond
backgrounds, where the energy of a level $n$ oscillator includes
an infinite number of $\alpha'$ corrections upon expanding the
square root $\sqrt{n^2+(\mu\alpha' p_+)^2}$: this can be traced to
the fact that perturbative strings are neutral under the Ramond
background, whereas here the NS background gives an additional
contribution to the mass.

If we now allow for a compact $u$ direction, the momentum $p_-$
becomes quantized in integer units (since the radius of $u$ is
fixed to be 1), and the spectral flows $w_L$ and $w_R$ can be
different integers. Let us denote $w=(w_L+w_R)/2$. The lhs of the
matching relation \eqref{match} receives an extra contribution \be
(w_L-w_R) \left( p_- - p_+ - w + \frac12 (1+n_L+n_R) \right ) \ee
allowing for more possibilities for level matching. If on the
other hand the coordinate $v$ is made compact, the momentum $p_+$
becomes quantized. In order to keep the conformal spin integer,
the radius of $v$ should be fixed to $2\pi$. It is then possible
to satisfy the matching relation \eqref{match} with $N_L \neq
N_R$. Finally, let us comment on type I states, with $p_+=0$.
Except for the tachyon, these states can carry only momentum along
$v$, that is along the wave itself. In fact, they can be thought
of as the constituents of the wave, in agreement with the fact
that $p_+=0$ encode the vacuum structure in light cone
quantization.

\subsection{The fate of discrete symmetries}

The pp-wave backgrounds $H_D$ threaded by NS flux \begin{equation}
\label{dspp1} ds^2=2dudv + \sum_{i=1}^{D\over 2} dw^id\bar w^i -\frac14 \mu^2
\left(\sum_{i=1}^{D/2} w^i\bar w^i\right) ~du^2\sp H=\mu ~du\wedge
\sum_{i=1}^{D/2} dw^i\wedge d\bar w^i
\end{equation}
have a geometrical discrete (permutation) symmetry $S_{D/2}$ that
permutes the $D/2$ 2-planes. This symmetry is acting on the $H_D$
symmetry algebra as an external automorphism, permuting pairs of
raising and lowering operators
\be P^{\pm}_{i}\leftrightarrow P^{\pm}_{j}
\ee

The ground-state of a generic type II/III representation
(corresponding to the harmonic oscillator ground state) can be
written as
\be
R^{\rm ground-state}_{II,III}=|\prod_{i=1}^{D/2}~H_{p_+}^i\rangle
\ee
where $H^i_{p_+}$ is the twist field of the i-th plane. Therefore
the $S_{D/2}$ symmetry is unbroken in the quantum theory. In the
special case $D=4$ corresponding to the Penrose limit of
$AdS_3\times S^3$  (near horizon region of NS5-F1 intersection)
the symmetry is $Z_2$ and it acts in a similar fashion as the
$Z_2$ that interchanges $AdS_5\leftrightarrow S^5$.

\section{Null holography}
Having the dynamics of particles and strings under control, we now turn to the
main motivation of this paper, namely to investigate whether a holographic
description analogous to AdS/CFT can hold in these homogeneous wave
backgrounds. A necessary condition is to set up a well defined Cauchy problem,
in order to relate boundary data on a holographic screen to bulk fields. This
is clearly not sufficient, e.g. in AdS one may have chosen time as a candidate
holographic direction, however one would have had to introduce two boundaries
at $t=\pm\infty$, as in the dS case.

In the pp-wave background \eqref{dspp}, the light-cone coordinate $u$ seems to
be the only Killing vector leading to a well-defined Cauchy problem. For
example, the oscillatory motion along $v$ precludes to use it as a time
coordinate, since fields at a given $v_0$ will propagate both forward and
backward. Another way to see that the propagation along $v$ is problematic is
that the kinetic term $\p_v^2$ in the wave equation \eqref{laphar} is singular
at $x_1^2+x_2^2=0$, so that only wave packets with no support at the bottom of
the potential may be propagated. Alternatively, one may be tempted to think of
the radial direction in transverse plane as the holographic direction, since,
in analogy with AdS, massive particles are confined along this direction.
However, this is also true of the massless ones (with the exception of the
$p_+=0$ modes, propagating strictly along the wave), so that this direction is
effectively compact, and not a boundary. On the other hand, the $u$ direction
is non-compact with smooth propagation and one can construct 
well-defined in and out-states.

Another crucial ingredient in the AdS/CFT correspondence is the relation
between holographic flow in the bulk and RG flow in the dual gauge theory (this
may not be a generic feature of holography, however we will conservatively
assume that this is true). This relation is particularly obvious in the
Poincar\'e patch, where the metric is the analytic continuation of the one on
the Poincar\'e upper half plane, \be ds^2= (-dt^2+dx_i^2 + dz^2)/z^2
\label{poin}\ee Going
to the boundary $z\to 0$ makes an infinite rescaling of the metric, hence
corresponds to the UV of the gauge theory, while the IR flow corresponds to
$z\to \infty$.

\subsection{Holography in the Poincar\'e patch}

Here we show that the pp-wave background can be put in a form very similar 
to the Poincar\'e patch of AdS \eqref{poin},
where the light-cone coordinate $u$ now generates the RG flow. Notice first
that like all the other maximally symmetric Lorentzian spaces, the pp-wave
background \eqref{dspp} is conformally flat. The conformal factor making the
curvature vanish is easily found to be $1/\sin^2((u-u_0)/2)$ for arbitrary
$u_0$. Each interval of length $2\pi$ can therefore be mapped through a
non-singular conformal transformation to Minkowski space.
The change of coordinates making the space manifestly conformal to Minkowski
space is easily found: starting from the coordinates in \eqref{dspp}, define
\be \label{chcf} x^+=-\cot(u/2)\ ,\quad v=x^- -\frac12(y_1^2+y_2^2)\sin u\
,\quad x_i=2y_i \sin(u/2) \ee The metric and flux then read \be \label{dscf}
ds^2=\frac{4}{1+x^{+2}}(dx^+ dx^- + dy^{2}_1+dy^{2}_2)\ ,\quad
H=\frac{1}{(1+x^{+2})^2} dx^+ dy_1 dy_2 \ee A motion along the $x^+$ direction
therefore appears to rescale the metric of a fixed $x^+$ slice. Our claim is
that there should exist {\it a holographic dual to string theory in a
pp-wave background of type $H_{D}$, with degrees of freedom
living on the degenerate space described by the  $(\xm,y_i)$ coordinates,
together with the other flat transverse coordinates}.
The correlators in the holographic dual are S-matrix elements relating the
incoming states at $\xp=-\infty$ to the outgoing states at $\xp=+\infty$. The
holographic direction is then the null direction $x^+$, to be contrasted with
the space-like (resp. time-like) direction in AdS (resp. dS).

This  claim requires some qualifications and comments.
Firstly, in contrast to AdS, the conformal factor $1/(1+\xp^2)$ never blows
up. It reaches a maximum at $\xp=0$, which implies that the UV is effectively
cut-off in the holographic dual. We view this as an indication that the dual
holographic degrees of freedom are non-local -- possibly a string theory of a
new type. On the other hand,  as $x^+\to \pm \infty$, the
conformal factor goes to zero, corresponding to the horizon of the
patch, so these regions should correspond to the IR of
the gauge theory. Attempts to define holography on the horizon are
not unprecedented \cite{Solodukhin:1998tc}, and this is also the
approach we follow here.

Next, propagation along a null coordinate $\xp$ involves a first-order
differential operator in $\xp$. This forbids to impose boundary conditions at
the two boundaries $\xp=\pm \infty$ at the same time, unlike the de Sitter
case. The dichotomy in AdS between non-normalizable solutions corresponding to
operator insertions and normalizable ones describing the vacuum
\cite{deHaro:2001xn} seems also to be lost.
Only the first kind is kept, since the
vacuum structure corresponds to $p_+=0$ states which cannot propagate along
$\xp$.
In fact, the Cauchy problem with a single slice at fixed $\xp$ is ill-defined:
as in Minkowski space, it is necessary to include boundary conditions on the
other side of the wedge, at fixed $\xm$. The vacuum structure should therefore
be specified on an extra ``holographic screen'' at fixed $\xm$.

Finally, the
conformal coordinate system only covers a patch $0<x^+<2\pi$ of the global
geometry \eqref{dspp} (or any translation thereof), which we may call
the Poincar\'e patch of the pp-wave geometry.
Even though we have seen that points at $u=0$ and $u=2\pi$ were an infinite
proper distance away unless $x_i=\pm x_i'$, this truncation may be
inconsistent. One would there have to go back to the global coordinates
$(u,v,x_i)$, where we still expect the coordinate $u$ to play the r\^ole of the
holographic direction.

\subsection{Geometry of conformal coordinates}

In order to evaluate the validity of this claim, let us first discuss
some aspects of the pp-wave geometry in these coordinates. To start with,
the Killing vectors corresponding to the action of $H_{4L} \times H_{4R}$
read
\bea
K&=&\p_- \nn\\
J_L&=&\frac{1+x^{+2}}{2}\p_+ -\frac{y_1^2+y_2^2}{2}\p_-
+ \frac12 (\xp y_1-y_2) \p_1 + \frac12 (\xp y_2+y_1) \p_2 \nn\\
J_R&=&\frac{1+x^{+2}}{2}\p_+ -\frac{y_1^2+y_2^2}{2}\p_-
+ \frac12 (\xp y_1+y_2) \p_1 + \frac12 (\xp y_2-y_1) \p_2\\
P_{iL}&=& y_i \p_- - \frac{\xp}{2} \p_i -\frac{1}{2}\epsilon_{ij} \p_j \nn\\
P_{iR}&=& y_i \p_- - \frac{\xp}{2} \p_i +\frac{1}{2}\epsilon_{ij} \nn \p_j
\label{sym}\eea where $\p_i:=\p/\p y_i$ and $\p_\pm:=\p/\p x^\pm$.
The translations along the three flat directions $(v,y_1,y_2)$ correspond to
the generators $K,P_{2L}-P_{2R}$ and $P_{1R}-P_{1L}$ respectively. The
Laplacian takes the form \be \Delta=\frac14
(1+x^{+2})\left(4\p_+\p_-+\p_1^2+\p_2^2\right)-\xp\p_- \ee where now
$\p_i=\p/\p y_i$. In particular, we see that the motion is not confined any
more in the transverse directions $y_1,y_2$.
The geodesics \eqref{geo} are in fact mapped to lines of arbitrary constant
transverse position, \be \xp(\tau)=-2\cot(p_+\tau)\ ,\quad \xm=p^2 \tau/p_+\
,\quad y_i=p_i/p_+ \ee
Remarkably, the slices of constant $\xp$ of the metric \eqref{dscf}
are invariant under all Killing vectors except for
\be
2P_-=J_L+J_R=(1+\xp^2)\p_+-(y_1^2+y_2^2)\p_-+\xp(y_1\p_1+y_2\p_2)
\label{hg}\ee This is the generator we want to identify with the
holographic direction: in terms of the original coordinates
\eqref{dspp}, it is simply the translation along the null $u$
direction. Assuming that the holographic dual indeed lives along
a slice of fixed $\xp$, we will have to find a realization of the
generator $P_-$ that does not involve the normal direction $\xp$:
we will come back to this point shortly.

\subsection{Generalized conformal wave background}

Having obtained the simple form \eqref{dscf} for the metric of the homogeneous
pp-wave in conformal coordinates, it is very natural to consider the more
general $D$-dimensional conformally flat (new) background with a null Killing
vector,
\be \label{dscfg} ds^2=\frac{1}{f^2(x^+)}(dx^+ dx^- + \sum dy_i^2)\
,\quad H=\sqrt{f''/f^{5}} ~dx^+ \wedge \omega
\ee
This solves the string
equations of motion, as long as the function $f$ is convex, $f''>0$ (as usual,
$\omega$ is a constant symplectic form on the $(D-2)$ dimensional transverse
space). The choice $f=\sqrt{1+\xp^2}/2$ corresponds to the homogeneous wave
background \eqref{dscf}.
The case where $f$ is constant
is of course flat Minkowski space. The case $f=\xp$
corresponds to the large $\xp$ limit of the homogeneous wave background, and is
also the same as flat space up to change of coordinates $(\xp,\xm,y^i)
\to(-1/\xp,\xm+ y_i^2/\xp,y^i/\xp)$. The background \eqref{dscf} is therefore
asymptotically flat at $x^+=\pm\infty$, and one may be able to define an
S-matrix relating the in-states at $x^+=-\infty$ to the out-states at
$x^+=+\infty$, after they interact with the wave around $|x^+|\sim 1$
\footnote{Note that one may also consider convex $f$-profiles that interpolate
between constant $f$ at $-\infty$ and linear at $+\infty$, or reverse.}.
A further generalization of the class \eqref{dscfg} would be to consider the
conformal rescaling of a Lorentzian space with a null Killing vector $\p_+$, by
a conformal factor depending on the null coordinate $\xp$.

The geodesics in the
generalized background \eqref{dscfg} are easily computed, and read \bea
y_i(\tau)&=& y_{i}+ \frac{p_i}{p_+}(\xp(\tau)-\xp) \nn\\
\xm(\tau)&=&\xm -\frac{p_i^2}{p_+^2}(\xp(\tau)-\xp)
+\frac{p_-}{p_+}\int_{\xp_0}^{\xp(\tau)} \frac{du}{f^2(u)}
\eea
where the proper time is related to the light-cone time
by $p_+ \tau= \int_{\xp}^\xp du/f^2(u)$.
The motion in the transverse directions $y^i$ is therefore free as expected. It
is interesting to note that when the integral $\int du/f^2$ converges, as in
the pp-wave case of interest, the motion is in fact massless at late times,
irrespective of the mass of the particle in the bulk. Another interesting case
is when $f(\xp)\to 1$ at $\pm\infty$: the coordinate $v$ then experiences a
translation by $(p_-/p_+)\int_{-\infty}^{\infty} (1/f^2-1)du$, as in standard
shock wave backgrounds.

\subsection{Null Cauchy problem}

If indeed $\xp$ is the holographic direction, one should be able to recover the
bulk configuration from ``initial'' data at the boundary $\xp=\xpo$ (eventually
taking the limit $\xpo\to -\infty$). We will now work out
the eigenmodes and propagator in the background \eqref{dscfg}.
More information and derivations are presented in appendix A.
Eigenmodes of the
Laplacian \be \Delta= f^2 (\Delta_T + 4 \p_+ \p_-) -(D-2) [f^2]'~\p_-
\label{lapcf} \ee (where $\Delta_T=\p_i\p_i$ is the Laplacian on the $(D-2)$
transverse coordinates) are easily computed for arbitrary $f$ profile, \be
\phi=f^{\frac{D-2}{2}}(\xp) \exp\left(i\left(p_+ x^- +p_i y_i -
\frac{p_i^2}{4p_+} \xp \right)-i \frac{m^2}{4p_+} \int_{-\infty}^{x^+}
\frac{du}{f^2(u)} \right)
\ee up to a normalization factor.
Notice that the phase is simply the one for a massless wave in Minkowski
space-time with momentum $p_-=- p_i^2/{4p_+}$, up to an additive phase shift
proportional to the integral of the conformal factor. The total phase shift
after crossing the curved background at $|x^+|\sim 1$ is \be \Delta\varphi =
\frac{m^2}{4p_+ \mu} \int_{-\infty}^\infty \frac{du}{f^2(u)} = \frac{\pi
m^2}{\mu p_+}\label{shift}\ee which we evaluated in the homogeneous pp-wave of
interest, after reinstating the mass scale $\mu$. This should be compared with
the phase shift in Minkowski space \be \frac{m^2}{4p_+ \mu}
\int_{-\infty}^\infty du = \frac{m^2\Delta \xp}{4\mu p_+} \ee which diverges
linearly.
The boundary to bulk propagator can be found by
adjusting the normalization so that $\phi(u,p_+,p_i)\to 1$ as $u\to u_0$. Going
back to position variables in transverse space, we get \bea
G(\xp,\xm,y_i;\xpo)&=& \int_{-\infty}^{\infty} {dp_-\over (2\pi)^{D-1}}
\left( \frac{4p_+ f(\xp)}{(\xp-\xpo)f(\xpo)} \right)^{\frac{D-2}{2}}\\
&& \exp\left(ip_+ \frac{\xm( \xp-\xpo) + y_i^2}{\xp-\xpo} -i \frac{m^2}{4p_+}
\int_{\xpo}^{\xp} \frac{du}{f^2(u)} \right)  \nn \eea The integral over $p_+$
is of the Bessel type, and yields \be G(\xp,\xm,y_i;\xpo)=
\theta(-r^2){(\xp-\xpo)^{{(2-D)\over 2}}\over (2\pi)^{D/2}i^{D-1}}
\left(\frac{f(\xp)}{f(\xpo)}\right)^{\frac{D-2}{2}} \left( \frac{\gamma
m^2}{r^2} \right)^\frac{D}{4} J_{D/2}\left( \sqrt{ {\gamma m^2 r^2\over
(\xp-\xpo)} } \right) \ee where $\theta(x)$ is the Heaviside function, $r^2$ is
the invariant distance from $(\xp,\xm,y_i)$ to $(\xpo,0,0)$ in Minkowski space,
$r^2=(x^+-x^+_0)\xm+y_1^2+y_2^2$, and $\gamma$ summarizes the effect of the
curved background, \be \gamma=\frac{1}{\xp-\xpo} \int_{\xpo}^{\xp}
\frac{du}{f^2(u)} \quad > 0 \ee
In particular for $f=1$ we recover the standard boundary to bulk propagator of
Minkowski space, $\gamma=\xp-\xpo$.  The main effect of the wave background is
therefore to stretch the invariant distance $r^2 \to \gamma r^2$,  i.e. to
introduce a refractive index $\gamma$. We also present the bulk propagator
which will be useful in holographic computations\be
G(\xp,\xm,y;{\xp}',{\xm}',y')=2^{(D-6)/2}[f(\xp)f(\xp')]^{D-2\over
2}~H(\xp-\xp')\ee
$$
\times \left({\pi^2m^2\gamma\over (\xp-\xpo)x^2}\right)^{D-2\over
4}~J_{D-2\over 2}\left( \sqrt{ {\gamma m^2 r^2\over (\xp-\xpo)} } \right) $$
%\vskip .7cm
Our derivation of the boundary to bulk propagator so far has been
casual and assumed special boundary conditions at $\xm \to \infty$. If order to
understand better the content of holographic correspondence in pp-wave
backgrounds, we need to allow for arbitrary states at $v=-\infty$. The careful
analysis of \cite{Tomaras:2001vs} in the context of two-dimensional QED on the
light cone can be transposed to our case as follows.
Let us consider a solution $\phi$ of $\Delta-m^2=0$, specified by its values on
the wedge $(\xp=\xpo,\xm>\xmo)$ and $ (\xm=\xmo,\xp>\xpo)$. Defining the
normalized field $\phi_-=\phi f^{(2-D)/2}$ and its canonical conjugate
$\phi_+=\p_- \phi_-$, the wave equation can be rewritten in the first order
form \be \p_- \phi_-=\phi_+\ ,\quad \quad 4 \p_+ \phi_+ + \Delta_T \phi_- =
{m^2\over f^2} \phi_- \label{a11}\ee
This can be integrated to (see appendix B)
\begin{equation}
\label{phip}
\hspace*{-2cm}
\begin{split}
\phi_+(\xp,\xm,y)=&\int_{-\infty}^{+\infty} \frac{dp_+}{2\pi}\left[
\int_{\xmo}^{+\infty} dv~e^{ip_+(v-\xm) }
\right. \exp\left(\frac{i}{4p_+} \int_{\xpo}^{\xp} \left(
\frac{m^2}{f^2(u)}-\Delta_T \right) du \right)
\phi_+(\xpo,v,y)\\
+{ie^{-ip_+(\xm-\xmo)}\over 4p_+}& \left. \int_{\xpo}^{\xp}
du~\left[\exp\left(\frac{i}{4p_+} \int_{u}^{\xp} \left(
\frac{m^2}{f^2(u')}-\Delta_T \right) du' \right)\right] \left(
\frac{m^2}{f^2(u)}-\Delta_T \right) \phi_-(u,\xmo,y) \right]
%$$
\end{split}
\end{equation}
where the integration prescription $p_+\to p_++i\epsilon$ is understood.
The conjugate field can be obtained as \be \hspace*{-2cm} \label{phim}
\begin{split}
&\phi_-(\xp,\xm,y)=\phi_-(\xp,\xmo,y) + \\
&+\int_{-\infty}^{+\infty}
\frac{dp_+}{2\pi}
\left[ \int_{\xmo}^{+\infty} dv~{i~e^{ip_+(v-\xm)
}\over p_+} \right.
\exp\left(\frac{i}{4p_+} \int_{\xpo}^{\xp} \left(
\frac{m^2}{f^2(u)}-\Delta_T \right) du \right)
\phi_+(\xpo,v,y)\\
&-{e^{-ip_+(\xm-\xmo)}\over 4p_+^2} \left.\int_{\xpo}^{\xp}
du~\left[\exp\left(\frac{i}{4p_+} \int_{u}^{\xp} \left(
\frac{m^2}{f^2(u')}-\Delta_T \right) du' \right)\right] \left(
\frac{m^2}{f^2(u)}-\Delta_T \right) \phi_-(u,\xmo,y) \right]
\end{split}
\ee
It can be verified that the solution above asymptotes properly at the limits
$\xp\to\xpo$ and $\xm\to\xmo$. The limit $\xp\to\xpo$ is fairly obvious from
the formula above. The other limit is subtle and the $i\e$ prescription is
crucial. Hence the values of $\phi_+$ at $\xp=\xpo$ and of $\phi_-$ at
$\xm=\xmo$ uniquely specify the field for all later $\xp$ and $\xm$.
\footnote{In Appendix B we show that these expressions
imply that there is no particle
production in this background when $p_+\not= 0$, as expected on general grounds
in backgrounds with a null Killing vector. This is in contrast to the constant
electric field case considered in \cite{Tomaras:2001vs}.}

Finally, let us consider the effect of taking the limit $\xm_0\to-\infty$
on the result \eqref{phip}.
Using \eqref{delta} we get
\be
\begin{split}
&\phi_+(\xp,p_+,y)= \exp\left(\frac{i}{4p_+} \int_{\xpo}^{\xp}
\left( \frac{m^2}{f^2(u)}-\Delta_T \right)du \right)
\phi_+(\xpo,p_+,y)+2\pi\delta(p_+)\times\\
&\times \int_{\xpo}^{\xp} du~\left[\exp\left(\frac{i}{4p_+} \int_{u}^{\xp}
\left( \frac{m^2}{f^2(u')}-\Delta_T \right) du' \right)\right] \left(
\frac{m^2}{f^2(u)}-\Delta_T \right) \phi_-(u,-\infty,y)
\end{split}
\label{phip2}
\ee
This shows that non-zero $\phi(x^+,\xm\to-\infty,y)$ generates delta function
singularities at $p_+=0$. Our discussion of holography will be restricted
for simplicity to the case where $\phi$ vanishes at $\xm\to-\infty$.
It would be interesting to develop techniques to regulate these contributions.

\subsection{Boundary symmetries and Ward identities}

At this stage, the boundary to bulk propagator provides us with a
very concrete way to set up the holographic correspondence: in
analogy with the $AdS_3$ case, we may think of the propagator
$G(\xp,\xm,y_i;\xpo,\xmo,y_{i0})
:=G(\xp,\xm-\xmo,y_i-y_{i0};\xpo)$ as the bulk closed string
vertex associated to a holographic operator localized at
$(\xmo,y_{i0})$ on the boundary. The action of $H_{4L}\times
H_{4R}$ on the bulk coordinate $(\xp,\xm,y_i)$ translates into an
action on the holographic coordinate $(\xmo,y_{i0})$.

As we have seen earlier, the solution to the Laplace equation in terms of the
boundary value at $\xp=\xpo$ and zero on the $\xmo=-\infty$ screen is
\be
\label{solution}
\begin{split}\phi_0&(\xp,\xm,y)=\int
dp_+~dp~\left({f(\xp)\over f(\xpo)}\right)^{D-2\over 2} \\
&\times \exp\left[i(p_+\xm+p\cdot y)-i{p^2\over 4p_+}(\xp-\xpo)-i{m^2\over
4p_+}\int_{\xpo}^{\xp}{du\over f^2(u)}\right]\phi(p_+,p,\xpo)
\end{split}
\ee

We focus on $D=4$ where $f=\sqrt{1+\xp^2}/2$. The Killing vectors ${\cal L}_i$
corresponding to the action of $H_{4L} \times H_{4R}$ in the bulk were given in
({\ref{sym}). Since they commute with the Laplacian, they can be commuted
through the boundary to bulk propagator and act on the boundary field
$\phi(p_+,p,\xpo)$. We define the boundary action of the symmetry (in momentum
space) as \be\begin{split} {\cal L}_i\phi_0&(\xp,\xm,y)=\int dp_+~dp~
\left({f(\xp)\over f(\xpo)}\right)^{D-2\over
2} \\
&\times \exp\left[i(p_+\xm+p\cdot y)-i{p^2\over
4p_+}(\xp-\xpo)-i{m^2\over 4p_+}\int_{\xpo}^{\xp}{du\over f^2(u)}\right]\hat
{\cal L}_i\phi(p_+,p,\xpo)
\end{split}\ee
{}From (\ref{sym},\ref{solution}) we obtain
\bea
\hat K&=&ip_+ \nn\\
\hat J_L&=&i{p_+\over 2}\Delta_p-{\xpo\over 2}(p_i\p_{p_i})+{1\over
2}\left(p_1\p_{p_2}-p_2\p_{p_1}\right)-i{p^2\over
8p_+}(1+{\xpo}^2)-i{m^2\over 2p_+} -{\xpo\over 2}\nn\\
\hat J_R&=&i{p_+\over 2}\Delta_p-{\xpo\over 2}(p_i\p_{p_i})-{1\over
2}\left(p_1\p_{p_2}-p_2\p_{p_1}\right)-i{p^2\over
8p_+}(1+{\xpo}^2)-i{m^2\over 2p_+} -{\xpo\over 2}\nn\\
%\eea \bea
\hat P_{iL}&=& -p_+\p_{p_i}-i{\xpo\over 2}p_i-{i\over
2}\epsilon^{ij}p_j \ ,\qquad \hat P_{iR}= -p_+\p_{p_i}-i{\xpo\over
2}p_i+{i\over 2}\epsilon^{ij}p_j \nn\label{sbou}\eea where
$\Delta_p=\p_{p_i}\p_{p_i}$.
It can be checked that ${\cal L}_i$ satisfy the conjugate algebra of $\hat
{\cal L}_i$. As expected, the generators are singular when $p_+=0$.
Interestingly, the dependence on the location $\xpo$ of the boundary can be
disposed of by performing the non-local field redefinition
\be \phi(p_+,p_1,p_2)=\tilde\phi(p_+,p_1,p_2)
\exp\left(-i \frac{p_1^2+p_2^2}{4p_+}\xpo\right)
\label{fr}\ee
The generators then take the simpler form \bea
\tilde K&=&ip_+ \nn\\
\tilde J_L&=&i{p_+\over 2}\Delta_p-\frac{i}{8p_+}(p_1^2+p_2^2+4m^2)
+\frac12 ( p_1 \p_{p_2}-p_2 \p_{p_1}) \\
\tilde J_R&=&i{p_+\over 2}\Delta_p-\frac{i}{8p_+}(p_1^2+p_2^2+4m^2)
-\frac12 ( p_1 \p_{p_2}-p_2 \p_{p_1}) \nn\\
\tilde P_{iL}&=& -p_+\p_{p_i}-{i\over 2}\epsilon^{ij}p_j \nn\ ,\qquad
\tilde P_{iR}= -p_+\p_{p_i}+{i\over 2}\epsilon^{ij}p_j \nn
\nn\label{sbou2}
\eea
Transforming back to position space using Fourier transform \be
\phi(\xpo,\xm,y)=\int dp_+dp~\exp\left[i(p_+\xm+p\cdot y)\right]~\tilde
\phi(p_+,p,\xpo) \ee we obtain the boundary representation of the $H_D\times
H_D$ symmetry, \bea\label{bounsym}
\tilde K&=&\p_- \nn\\
\tilde J_L+\tilde J_R &=& -y_i^2 \p_- +{1\over 4}(4m^2-\Delta)\p_-^{-1}\\
\tilde J_L-\tilde J_R &=& y_1 \p_2-y_2 \p_1 \nn \\
\tilde P_{iL}&=& y_i \p_-  -\frac{1}{2}\epsilon_{ij} \p_j\ ,\qquad
\tilde P_{iR}= y_i \p_-  +\frac{1}{2}\epsilon_{ij} \p_j
\nn\eea
where $\p_-^{-1}=\int d\xm$, $\p_-^{-1}\p_-=1$. This result
can be generalized to arbitrary even $D$ by replacing $J_L+J_R$ by
the rotation in all 2-planes simultaneously.

The formulae \eqref{bounsym} thus describe the action of the bulk
isometries on the boundary data. They correspond to symmetries
that must be satisfied by the correlation functions
of the putative dual gauge theory. They suggest the existence of
a novel class of (possibly non-local) field theories where
conformal invariance is replaced by
invariance under the (super) Heisenberg-type
group $H_D^L\times H_D^R$. The analogue of the conformal dimension
is the mass $m^2$, which is also the Casimir of this representation.
The holographic generator is the combination $J_L+J_R$
which is nothing but the Hamiltonian of the harmonic oscillator
with frequency $|\p_-|/2$. Note in particular that the rescaling
operator appearing in the last term of the bulk generator \eqref{hg}
has disappeared in the process of going to the boundary, due to
the non-local field redefinition \eqref{fr}. In terms of the
new field $\tilde \phi$, the holographic generator does not appear
to involve a RG flow any more. Alternatively, one may have located
the holographic screen at $x_0^+=0$, i.e. as close to the UV regime
as one can gets, so that the field redefinition  \eqref{fr} would
not be necessary. Since the conformal factor attains a maximum at
$\xp=0$, one recovers the conclusion that the holographic generator
should not involve any rescaling of the transverse coordinates.

It is interesting to determine the consequences of the
Ward identities associated to this symmetry on the $n$-point
functions correlators.
Invariance under the generators $K$ and $P_{iL}-P_{iR}$ enforces momentum
conservation in all directions $\xm,y_i$. For the two-point function,
invariance under $P_{iL}+P_{iR}$ then implies that the correlator is
independent of the transverse momenta $p_i$. Invariance under $J_L,J_R$
requires the masses of the two fields to be the same. The most general
two-point function consistent with the symmetry is therefore \be \langle
\phi_{m_1}(p_+,p)\phi_{m_2}(q_+,q)\rangle
=f_2(p_+)\delta(p_++q_+)\delta^{(2)}(p+q) \delta_{m_1,m_2}
\label{f2pt}\ee
where $f_2(p_+)$
is an arbitrary function of $p_+$.
Of course, this discussion cannot rule out contact
contributions at $p_+=0$, where the symmetry is ill-defined. The constraints
on the 3-point function are further analyzed in Appendix D.

\subsection{Two-point and three-point functions}

Let us now try and derive the boundary correlators following the same logic as
in the AdS/CFT case. There, the prescription was that the classical action of
the solution of the bulk equations of motion satisfying appropriate boundary
conditions, gave the generating functional for the off-shell correlation
functions of the boundary operators \cite{Gubser:1998bc}. We will follow the
same logic here, and present a sample computation for a self-interacting
massive scalar field in the bulk in order to illustrate our methodology.
Of course, a full-fledged computation should involve the whole supergravity
multiplet in the bulk, or rather the whole tower of string states, which
we will leave for future work. As discussed below \eqref{phip2}, we will also
for simplicity assume zero boundary conditions on the holographic screen
at $\xm\to -\infty$.
These are the only boundary conditions that leave the $H_{4L}
\times H_{4R}$ symmetry
unbroken in the boundary theory.
It would be interesting to extend the computation
below to the more general case, in order to study the effect on
the correlators of changing the vacuum.

Let us therefore
consider a self-interacting minimally-coupled massive scalar field in
the bulk, with action \be S=-\frac12 \int d^{D}x \sqrt{-g} \left( g^{\mu\nu}
\p_\mu \phi ~ \p_\nu \phi +m^2 \phi^2 +{2\lambda\over 3}\phi^3\right) \ee The
equations of motion $(\Delta-m^2)\phi=\l \phi^2$ can be solved perturbatively
in $\l$ by expanding $\phi=\phi_0+\l \phi_1+{\cal O}(\l^2)$ to obtain \be
(\Delta-m^2)\phi_0=0\sp (\Delta-m^2)\phi_1=\phi_0^2 \ee
We now evaluate the
action $S=S_0+\l S_1+{\cal O}(\l^2)$ on the classical solution in order to
obtain the boundary correlators. This amounts to computing S-matrix
transition amplitudes between $\xp=-\infty$ and $\xp=+\infty$.
As usual, the on-shell action
reduces to boundary terms,
\bea S_0&=&-\frac12 \int d^{D}x
\sqrt{-g} \left( g^{\mu\nu}
\p_\mu \phi_0 ~ \p_\nu \phi_0 +m^2 \phi_0^2\right)\\
&=&-\int dx^+~ dx^-~ d^{D-2}y ~ \left[
\p_+ \left(f^{2-D} \phi_0 \p_- \phi_0 \right)+ \p_- \left(f^{2-D} \phi_0 \p_+
\phi_0 \right)\right]\nn\\
S_1&=&-\l\int d^{D}x \sqrt{-g} \left( g^{\mu\nu} \p_\mu \phi_0 ~ \p_\nu \phi_1
+m^2 \phi_0\phi_1 +{1\over 3}\phi_0^3\right)  \\
&=& -{2\l\over 3}\int \left[2\p_+\p_-(f^{2-D} \phi_0 \phi_1)+\p_+ \left(f^{2-D}
\phi_1 \p_- \phi_0 \right)+\p_- \left(f^{2-D} \phi_1 \p_+
\phi_0\right)\right]\nn
\eea
This is evaluated in detail in Appendix C.
Remarkably, the first order correction $S_0$
vanishes, implying a vanishing two-point function for the boundary field
$\phi$. As we explain in Appendix D, this is consistent with the Ward
identities, although the latter would allow a more general 2-point function.
The three-point amplitude is however non-trivial. It can be rendered finite by
a wave function renormalization $\phi_0\to \phi_0~f(\xpo)^{2-d\over 2} $.
After amputating the renormalized sources and momentum conserving
$\delta$-functions,
the three-point amplitude is given by
\be \label{3point}F_3(p,q,r) =\l {(2\pi)^{D-1}\over
12}\int_{\xpo\to -\infty}^{\infty}du~f(u)^{D-6\over 2}\exp\left[-{im^2\over
4}\left({1\over p_+}+{1\over q_+}+{1\over
r_+}\right)\tilde\gamma(u,\xpo)\right] \ee
$$\times
\exp\left[-{i\over 4}\left({p^2\over p_+}+{q^2\over q_+}+{r^2\over
r_+}\right)(u-\xpo)\right]
$$
with \be \tilde\gamma(x^+,\xpo)\equiv \int_{\xpo}^{\xp}{du\over
f^2(u)} \ee It can be checked that this result is compatible with
the Ward identity \eqref{f3pt}. Extending the domain of
integration to the whole space-time (i.e. integrating from
$u=-\infty$ to $u=+\infty$ in global coordinates), the same result
is obtained, multiplied by an extra delta function enforcing
conservation of $m^2/p_+$ modulo integers. This is in agreement
with the conservation of $p_-$ in global coordinates. The
computation of higher amplitudes is in principle straightforward.
It would be interesting to understand the effect of other boundary
conditions at $\xm\to-\infty$ on these correlators.

\section{Summary, conclusions and open problems}

In this paper we have studied aspects of string propagation and
holography homogeneous pp-wave backgrounds.
These are exact string embeddings of the maximally
symmetric Cahen-Wallach spaces which, in the case where there
are supported by Neveu-Schwarz flux, can also be thought of as the group
manifold of the extended Heisenberg group $H_D$. They arise as Penrose limits
of the near horizon region of the the NS5-brane (blown up around a geodesic
spinning around $S^3$) (for $H_4$), $AdS_3\times S_3$ ($H_6$) as well as higher
intersections of two NS5-branes with fundamental strings ($H_8,H_{10}$).
In addition, they are interesting tractable exemples of time-dependent
backgrounds in string theory.

In the first part of this paper, we have discussed aspects of string
propagation in these backgrounds, elaborating on earlier studies
\cite{Kiritsis:jk,Kiritsis:1994ij}. The study of the geodesics and solutions of
the wave equation has shown that the natural time variable is the null
coordinate $\xp$. An observer co-moving with the wave has $\xp=$constant. Time
slices with $\Delta\xp=2\pi \Zint$ have been shown to be an infinite proper
distance away from each other, except for coinciding positions in transverse
space.

We have subsequently discussed the covariant quantization of the NSR string in
these backgrounds, using in particular the free-field resolution of the current
algebra which enables us to construct vertex operators for all the affine
representations. Type I states correspond to modes with $p_+=0$, travelling
with the wave. They would be invisible in light-cone quantization. Their vertex
operators are simple exponentials. They are massless and interact among
themselves as in flat space, leading to only type I states in intermediate
channels.

Type II and III states are localized at the origin in the transverse $D-2$
dimensional space, and their vertex operators are constructed in terms of twist
fields of free complex bosons by an angle $2\pi p_+$, in general irrational.
They have non-trivial interactions, and it is an open problem to calculate
their S-matrix. Solving the associated Knizhnik-Zamolodchikov equations is the
most direct approach to this problem.

Since twist fields only exist for $0<1< p_+$, the infinite range of $p_+$ is
generated by the spectral flow(s) of the $H_D$ current algebra. The new states
are associated with long strings winding (and oscillating under the influence
of the wave) in the transverse harmonic potential. At integer values of $p_+$,
they are also long strings that expand linearly with light-cone time in the
transverse directions. For specific value of the energy, they exist as static
circular strings, and can change their radius at no cost of energy. The
inclusion of spectral flow restores an affine-Weyl symmetry  and should be
responsible for improved properties of the theory under T-duality
\cite{Kiritsis:1991zt}.
Moreover, discrete symmetries analogous to the $Z_2$ symmetry
$AdS_5\leftrightarrow
S^5$ arise here as well and we have shown that they are symmetries
 of the quantum theory.

It would be interesting to better understand the similarities of the spectrum
with the $AdS_3$ case. In this respect, it may be useful to remember that the
generators of $Sl(2)$ can be constructed in the enveloping algebra of the
harmonic oscillator.

We should mention that the vacuum amplitude (this term is more appropriate than
partition function since there is no Euclidean continuation) was computed in
\cite{Kiritsis:1994ij} and found to be equal to that of flat space. There are
two potentially non-trivial pieces in this amplitude: the contribution of the
zero-level of the current algebra and the existence of null states at
$p_+\in\Zint$. The second is eventually irrelevant since such contributions are
of measure zero and $p_+$ is continuous. In fact this is also true in the case
of strings in flat Minkowski space-time. At special values of the momenta,
there are extra physical states associated to existence of null vectors in the
Verma modules. However, they are not visible in the vacuum amplitude.

 Summing
over the zero-level states (the infinite dimensional reps of $H_{4L}
\times H_{4R}$)
gives infinity since they are all degenerate in energy. This is a volume
divergence and is analogous to summing over primaries in Minkowski space in a
radial coordinate representation of the spectrum.
The unambiguous way to sum
them is to compute the supergravity finite-time propagator at coincident points
(which is also presented in  appendix A) and then integrate over the volume.

The exact CFT solution of CW spaces can be written in terms of operators
occurring in standard orbifolds of flat space, but for irrational twists. This
intriguing fact seems to be related to the observation that such backgrounds
are related by formal T-dualities to flat space \cite{Kiritsis:jk,dual},
presumably with identifications. Clarifying this fact should give a better
understanding of the free-field resolution of the affine algebra and have
further applications. It is conceivable that this remains true also for CW
spaces supported by RR backgrounds (relevant for the Penrose limits of
near-geometries of D-branes).

One of the main motivations for this study
was to try and understand holography in this
type of gravitational backgrounds. We have argued that the appropriate patch to
carry out this analysis was the analogue of the Poincar\'e patch of AdS, where
the metric is conformal to flat Minkowski space. This patch covers
an interval of length $2\pi$ along the light-cone direction $u$,
and the conformal factor depends on $u$ only. The natural
holographic direction is then the null light-cone time $\xp=-2\cot(u)$. We have
proposed that holography relates S-matrix elements between two holographic
screens at $\xp\to\pm \infty$, to off-shell correlation functions in a
putative gauge theory dual living on a slice at fixed $\xp\to-\infty$.

The isometries $H_{DL} \times H_{DR}$ of the bulk are realized as
a Schr\"odinger-like representation of the extended Heisenberg algebra on the
boundary, whose Ward identities restrict the form of the correlators. We have
provided a sample computation of the two- and three-point functions of a
massive bulk scalar field. The two-point function seems to vanish while the
three-point functions has a non-trivial structure. These quantities should
eventually be calculable in string theory. It would be very
interesting to have a microscopic definition of the dual theory, to compare our
results with. It is unclear at this stage how to construct
field theories with the required Heisenberg symmetry.

In contrast to the space-like or time-like holographic
set-ups in AdS or dS, the evolution along a null direction is first order in
time, which implies that there is a single set of modes at a given mass. These
modes are the analogues of the non-normalizable modes in AdS, corresponding to
deformations of the boundary theory.
In fact, a careful analysis of the Cauchy problem shows that a complete
specification of the field in the bulk requires boundary conditions both at
$\xp=-\infty$ and $\xm=-\infty$. Non-trivial boundary values at $\xm=-\infty$
give contributions localized at $p_+=0$ and are thus associated with the vacuum
structure. We have chosen the simplest such boundary conditions (vanishing
field at $\xm=-\infty$) for  our computations. Non-trivial boundary conditions
here induce divergences at $p_+=0$ and it is an open problem to deal with
them.

We believe that a better understanding of such backgrounds will shed light on
non-space-like versions of holography, and hopefully on holography in flat
Minkowski space-time. In addition, it will be interesting to pursue their
study from the point of view of time-dependent backgrounds, as they
offer interesting solvable exemples of cosmological universes, where
the r\^ole of time is played by a null coordinate.

\enlargethispage{3mm}

\acknowledgments The authors are grateful to Th. Damour, G. d'Appollonio, J. de
Boer, J. Gomis, N. Itzhaki, C. Kounnas, N. Nekrasov, G. Papadopoulos and C.
Schweigert for useful discussions. E. K. acknowledges the hospitality of Centre
Hospitalier d'Orsay where part of this work was carried out. B. P. thanks
the Harvard and U Penn theory groups for hospitality and discussions
subsequent to the release of the first draft of this paper. The work of E. K.
was partially supported by Marie Curie contract MCFI-2001-0214 and RTN
contracts HPRN--CT--2000--00122 and --00131. \vskip 1.5cm

\vskip 5mm
{\it \noindent Note added.}
After this article was released, a number of works appeared discussing
holography in pp-wave backgrounds from different viewpoints and reaching
different conclusions. In \cite{Leigh:2002pt}, by following the AdS/CFT
correspondence through
the Penrose limit, the holographic direction was argued to be the radial
distance in the transverse directions $\Real^{p-1}$
originating from the AdS factor in $AdS_{p}\times S^p$. In this
approach, the holographic degrees of freedom still live in $(p-1)$-dimensional
Minkovski space as in the AdS parent, even though the boundary of AdS
decouples in the Penrose limit.
On the other hand, in \cite{Berenstein:2002sa},
the conformal infinity of the pp-wave
background is identified as a null trajectory moving along $u$ at infinity
in the $(v,x_i)$ space. Based on this, a holographic description
as a quantum mechanical model was suggested,
in analogy with the M(atrix) model
\cite{Banks:1996vh}.
These different approaches do not necessarily
exclude each other, as there may be different holographic
descriptions of the same geometry, differing by a possibly complicated
change of variables. While all these approaches await a concrete
definition of the dual theory, ours makes the symmetries manifest and
allows for a direct boundary to
bulk correspondence, which we believe is crucial in implementing
holography.

\appendix
\vskip 10mm

\centerline{\it\bf Appendices}

\section{Propagators}

In this appendix we will derive and study bulk-to-bulk and bulk-to-boundary
propagators for global and conformal coordinates and study their properties. We
will choose boundary conditions such that fields vanish at $v=-\infty$ in
global coordinates and at $\xm=-\infty$ for conformal coordinates.

\subsection{Boundary and bulk to bulk propagators in global coordinates} The
global coordinates were defined in \eqref{dspp}. For simplicity we will confine
ourselves to $H_4$ (D=4). At fixed $p_+$, the wave equation in global
coordinates is equivalent to the Schr\"odinger equation for a two-dimensional
harmonic oscillator of frequency $|p_+|/2$. We will use the propagator for a
single harmonic oscillator of frequency $\omega$ and mass $m$ \bea
G(x_1,x_2;t)&\equiv& \sum_n
\psi_n^*(x_1)\psi_n(x_2)e^{-iE_nt} \\
&=&\sqrt{im\omega\over 2\pi \sin
\omega t}\exp\left[{im\omega\over \sin\omega t}\left({1\over
2}(x_1^2+x_2^2)\cos\omega t -x_1x_2\right)\right]
\eea
normalized so that
\be \lim_{t\to 0}G(x_1,x_2;t)=\delta(x_1-x_2)\ee
Let $\lambda$ be
the eigenvalues of the Laplacian \be \Delta F_{\lambda}=\lambda
F_{\lambda}\sp \lambda=-p^2-2p_+p_--|p_+|(n_1+n_2+1)\ee with \be
F_{\lambda}=\exp\left[ i (p_+ v + p_- u + p \cdot
y)\right]\psi_{n_1}(x_1)\psi_{n_2}(x_2) \ee
For a massless mode, the propagator in Schwinger time is therefore
\be
\begin{split}
&G(u,v,\vec y,\vec x;u',v',\vec y',\vec
x';t)\equiv\langle u,v,\vec y,\vec x|e^{it\Delta}|u',v',\vec
y',\vec x\rangle \\
&={1\over (2\pi)^{d+2}}\sum_{n_{1,2}}\int dp_+~dp_-~d^d p~
\exp[-ip_+(v-v')-ip_-(u-u')-ip\cdot(y-y')]\\
&\times\exp[-it(p^2+2p_+p_-+|p_+|(n_1+n_2+1))]\psi^*_{n_1,n_2}(x)
\psi_{n_1,n_2}(x')\\
&={i\over (2\pi)^{d+2}}{1\over 8t^2}\left({\pi\over
it}\right)^{d/2}{u-u'\over
\sin[(u-u')/4]}\exp\left[{i(y-y')^2\over 4t}\right]
\exp\left[{i(u-u')(v-v')\over 2t}\right]\times \\
&\times\exp\left[ {i(u-u')\over 4t\sin[(u-u')/4]}\left( {1\over
2}(x_1^2+x_2^2+x_1'^2+x_2'^2)\cos[(u-u')/4]-x_1x_1'-x_2x_2'\right)\right]
\end{split}
\ee
where $d$ is the number of transverse flat directions ($d=6$ for the NS5
problem). It satisfies \be \lim_{t\to 0}G(u,v,\vec y,\vec x;u',v',\vec y',\vec
x';t)=\delta(u-u')\delta(v-v')\delta^{(2)}(x-x')\delta^{(d)}(y-y') \ee In the
coincidence limit, one obtains \be G(u,v,\vec y,\vec x;u,v,\vec y,\vec
x;t)={1\over 2^{d+3}\pi^{d/2+2}i^{d/2-1}t^{2+d/2}} \ee This value is equal to
that of flat (d+4)-dimensional flat space.
 This was used in
\cite{Kiritsis:1994ij} in order to compute the vacuum amplitude and show that
it is equal to that flat space.
The massive bulk propagator is defined as \be (\Delta-m^2)
H=-\delta(u-u')\delta(v-v')\delta^{(2)}(x-x')\delta^{(d)}(y-y')\ee and can be
computed from $
H(z,z')=-\sum_{\lambda}F^*(z)_{\lambda}F(z')_{\lambda}/(\lambda-m^2)$. We
obtain \be
\begin{split}
H=&\sum_{n_1,n_2}\int~{dp_+dp_-d^d p\over (2\pi)^{d+2}}
exp[-ip_+(v-v')-ip_-(u-u')-ip\cdot(y-y')]  \\
&\times {\psi^*(x_1,x_2)_{n_1,n_2}\psi(x_1',x_2')_{n_1,n_2}\over
p^2+2p_+p_-+|p_+|(n_1+n_2+1)+m^2}
\end{split}
\ee
We perform the contour integral in $p_-$ and use the oscillator propagator to
obtain \be\begin{split} &H={\pi\over 4}\int {dp_+d^d p\over
(2\pi)^{d+2}\sin[(u-u')/2]}
\exp\left[-ip_+(v-v')-ip\cdot(y-y')+i{p^2+m^2)\over 2p_+}(u-u')\right]\\
&\times \exp\left[{i|p_+|\over 2\sin[(u-u')/2]}\left({1\over
2}(x_1^2+x_2^2+x_1'^2+x_2'^2)\cos[(u-u')/2]-x_1x_1'-x_2x_2'\right)\right]\\
&=\left(2\pi\over -i(u-u')\right)^{d/2}\int {p_+^{d/2}dp_+\over
8(2\pi)^{d+1}\sin[(u-u')/2]}\exp\left[-ip_+\left(v-v'+{1\over 2}{(y-y')^2\over
u-u'}\right)+i{m^2(u-u')\over 2p_+}\right]\\
&\times\exp\left[{i|p_+|\over 2\sin[(u-u')/2]}\left({1\over
2}(x_1^2+x_2^2+x_1'^2+x_2'^2)\cos[(u-u')/2]-x_1x_1'-x_2x_2'\right)\right]
\end{split}\ee
and we finally obtain
\be
H={1\over 4i^{d-1}(2\pi)^{(d+2)/2}}{u-u'\over
\sin[(u-u')/2]}\left({m\over r}\right)^{d+2\over 2}K_{d+2\over
2}(mr)
\ee
where $r$ is the invariant geodesic distance given in (\ref{distance}).

\subsection{Propagators in conformal coordinates}

Here we work in conformal coordinates \be  ds^2=\frac{1}{f^2(x^+)}(dx^+ dx^- +
\sum dy_i^2)\ ,\quad H=\sqrt{f''/f^{5}} ~dx^+ \wedge \omega \ee The eigenmodes
of the Laplacian \be \Delta= f^2 (\Delta_T + 4 \p_+ \p_-) -(D-2) [f^2]'~\p_-
\label{lapla} \ee (where $\Delta_T=\p_i\p_i$ is the Laplacian on the $(D-2)$
transverse coordinates) are easily computed for arbitrary $f$ profile, \be
\phi=f^{\frac{D-2}{2}}(\xp) \exp\left(i\left(p_+ x^- +p_i y_i -
\frac{p_i^2}{4p_+} \xp \right)-i \frac{m^2}{4p_+} \int_{-\infty}^{x^+}
\frac{du}{f^2(u)} \right) 
\ee
We will solve the massive scalar equation $(\Delta-m^2)\phi=0$ and allow
for an arbitrary dimension $D$ for the CW space. Additional flat coordinates
can be accommodated by substituting $m^2\to m^2+P^2$. We first write the
solution in terms of the boundary value (at $\xp=\xp_0$) in momentum space. We
pick also the boundary condition $\phi(\xp,-\infty,y)=0$. From now on the
integration ranges of $p_+$ and the $D-2$ transverse momenta $p$ are the whole
real line: \be\begin{split} \phi_0(\xp,\xm,x)=&\int
dp_+~dp~\chi(p_+,p)\left({f(\xp)\over f(\xpo)}\right)^{D-2\over 2} \\
&\times\exp\left[i(p_+\xm+p\cdot x)-i{p^2\over 4p_+}(\xp-\xpo)-i{m^2\over
4p_+}\int_{\xpo}^{\xp}{du\over f^2(u)}\right]
\end{split}\ee
where $\chi)p_+,p)$ is the Fourier transform of the boundary data at
$\xp=\xpo$, \be \lim_{\xp\to\xpo}\phi_0(\xp,\xm,x)=\int
dp_+~dp~\chi(p_+,p)\exp\left[i(p_+\xm+p\cdot x)\right]\equiv \chi(\xm,x) 
\ee
{}From this we can obtain the bulk-to-boundary propagator \be
K(\xp,\xpo;\xm,x;\xm',x')=\int {dp_+~dp\over (2\pi)^{D-1}}~\left({f(\xp)\over
f(\xpo)}\right)^{D-2\over 2} \ee $$\times \exp\left[i(p_+(\xm-\xm')+p\cdot
(x-x'))-i{p^2\over 4p_+}(\xp-\xpo)-i{m^2\over 4p_+}\int_{\xpo}^{\xp}{du\over
f^2(u)}\right]
$$
normalized so that  \be
\lim_{\xp\to\xpo}K(\xp,\xpo;\xm,x;\xm',x')=\delta(\xm-\xm')\delta^{(D-2)}(x-x')
\ee

We can perform the momentum integrals in order to obtain the propagator in
configuration space, \be G(\xp,\xm,y_i;\xpo)=
\theta(-r^2){(\xp-\xpo)^{{(2-D)\over 2}}\over (2\pi)^{D/2}i^{D-1}}
\left(\frac{f(\xp)}{f(\xpo)}\right)^{\frac{D-2}{2}} \left( \frac{\gamma
m^2}{r^2} \right)^\frac{D}{4} J_{D/2}\left( \sqrt{ \gamma m^2 r^2} \right) \ee
where $\theta(x)$ is the Heaviside function, $r^2$ is the invariant distance
from $(\xp,\xm,y_i)$ to $(\xpo,0,0)$ in Minkowski space,
$r^2=(x^+-x^+_0)\xm+y_1^2+y_2^2$, and $\gamma$ summarizes the effect of the
curved background, \be \gamma=\frac{1}{\xp-\xpo} \int_{\xpo}^{\xp}
\frac{du}{f^2(u)} \quad > 0 \ee

We would now like to compute the bulk propagator satisfying \be (\Delta
-m^2)G(\xp,\xm,x;{\xp}',{\xm}',x')={1\over
\sqrt{g}}\delta(\xp-{\xp}')\delta(\xm-{\xm}')\delta^{(D-2)}(x-x') 
\ee 
Fourier
transforming
\be G(\xp,\xm,x;{\xp}',{\xm}',x')=\int
dp_+~dp~e^{i(p_+(\xm-{\xm'})+ p\cdot (x-x')}G(\xp,{\xp}',p_+,p) \ee 
we obtain
\be 4ip_+f^2(\xp)\p_{+}G-[p^2f^2(\xp)+m^2+i(D-2)p_+(f^2(\xp))']G
={f^D(\xp)\over (2\pi)^{D-1}}\delta(\xp-{\xp}') \ee

The solution of the differential equation above can be written as  \be
G(\xp,\xp',p_+,p)= {[f(\xp)f(\xp')]^{D-2\over 2}\over
(2\pi)^{D-1}8ip_+}H(\xp-\xp') \exp\left[-{ip^2\over 4p_+}(\xp-\xp')-{im^2\over
4p_+}\int_{\xp'}^{\xp}{du\over f^2(u)}\right] \ee where $H(x)=
c\theta(x)+(c-2)\theta(-x)$. $c=0,2$ corresponds to advanced or retarded
propagators. In order to treat symmetrically particles and anti-particles, we
choose $c=1$.
Performing the momentum integrals we obtain \be\begin{split}
G(\xp,\xm,x;{\xp}',{\xm}',x')=&\\
{2^{(D-6)/2}\over (2\pi)^{D-1}}[f(\xp)f(\xp')]^{D-2\over 2} & H(\xp-\xp')
\times \left({\pi^2m^2\gamma\over (\xp-\xpo)x^2}\right)^{D-2\over
4}~J_{D-2\over 2}\left( \sqrt{ \gamma m^2 r^2 } \right) \end{split}\ee

\section{The general scalar solution to the wave equation}
In this appendix, we derive the solution $\phi$ of the wave
equation $\Delta-m^2=0$ in the generalized conformal background
\eqref{dscf}, specified by its values on
the wedge $(\xp=\xpo,\xm>\xmo)$ and $ (\xm=\xmo,\xp>\xpo)$. Defining the
normalized field $\phi_-=\phi f^{(2-D)/2}$ and its canonical conjugate
$\phi_+=\p_- \phi_-$, the wave equation can be rewritten in the first order
form \be \p_- \phi_-=\phi_+\ ,\quad \quad 4 \p_+ \phi_+ + \Delta_T \phi_- =
{m^2\over f^2} \phi_- \label{a1}\ee This can be integrated to
 \bea
\phi_-(\xp,\xm,y)&=&\phi_-(\xmo,\xp,y)+\int_{\xmo}^{\xm} dv \phi_+(\xp,v,y) \\
\phi_+(\xp,\xm,y)&=&\phi_+(\xpo,\xm,y)+{1\over 4}\int_{\xpo}^{\xp}
du \left( \frac{m^2}{f^2(u)}-\Delta_T \right) \phi_-(u,\xm,y) \eea
where $y$ denote collectively the $D-2$ transverse coordinates.
Iterating these equations, we get a series expansion for $\phi_+$,
\bea &&\phi_+(\xp,\xm,y)=\sum_{n=0}^{\infty}
\int_{\xpo}^{\xp}\!\!\!\! du_1 \int_{\xpo}^{u_1}\!\!\!\!  du_2
\dots\int_{\xpo}^{u_{n-1}}\!\!\!\!  du_n \int_{\xmo}^{\xm}\!\!\!\!
dv_1 \int_{\xmo}^{v_1} \!\!\!\! dv_2 \dots\int_{\xmo}^{v_{n-1}}
\!\!\!\!  dv_n
\\&\times& \prod_{i=1}^n {1\over 4}\left( \frac{m^2}{f^2(u_i)}-\Delta_T \right) \left[
\phi_+(\xpo,v_n,y)+{1\over 4}\int_{\xpo}^{u_n} du \left(
\frac{m^2}{f^2(u)}-\Delta_T \right) \phi_-(u,\xmo,y) \right] \nn
\eea We now extend the integration range of $v_n$ to $+\infty$ by
using the integral representation \be
\theta(v_{n-1}-v_n)=\int_{-\infty}^{+\infty} \frac{dp_+}{2\pi}
\frac{i e^{-i(p_++i\epsilon)(v_{n-1}-v_n)}}{p_+ + i\epsilon}
\label{theta}\ee where $\epsilon$ is a positive number unspecified
at this stage. Renaming $v=v_{n-1}$ and integrating successively
along $v_{n-1},v_{n-2}\dots v_1$ yields \bea
\phi_+(\xp,\xm,y)&=&\int_{-\infty}^{+\infty} \frac{dp_+}{2\pi}
\int_{\xmo}^{+\infty} dv~ e^{i(p_++i\epsilon)(v-\xm)}
\sum_{n=0}^{\infty} \int_{\xpo}^{\xp}\!\!\!\! du_1
\int_{\xpo}^{u_1}\!\!\!\!  du_2
\dots\int_{\xpo}^{u_{n-1}}\!\!\!\!  du_n \\
\left(\frac{i}{p_+ + i\epsilon}\right)^n &\times& \prod_{i=1}^n {1\over
4}\left( \frac{m^2}{f^2(u_i)}-\Delta_T \right) \left[ \phi_+(\xpo,v,y)+{1\over
4} \int_{\xpo}^{u_n} du \left( \frac{m^2}{f^2(u)}-\Delta_T \right)
\phi_-(u,\xmo,y) \right] \nn \eea Using the identities \bea \int_0^{x^+} du_1
f(u_1) \int_0^{u_1} du_2 f(u_2) \cdots \int_0^{u_{n-1}}du_n f(u_n) &=& {1 \over
n!}
\left[\int_0^{x^+} dy f(y)\right]^n \\
\int_0^{x^+} du_1 f(u_1) \cdots
\int_0^{u_{n-1}} du_n f(u_n)
\int_0^{u_n} du g(u)
&=& \int_0^{x^+} du g(u) {1\over{n!}} \left[ \int_u^{x^+} dy f(y)\right]^n\nn
\eea
one easily obtains the expression \eqref{phip}. The conjugate field
$\phi_-(\xp,\xm,y)$ is then obtained by inverting \eqref{a1}, \be\phi_-=(4 \p_+
\phi_+)/\left({m^2\over f^2}- \Delta_T\right)\ee leading to \eqref{phim} for
$\phi_-=f^{(2-D)\over 2}\phi$. Having reached this stage of the computation, it
takes little more effort to investigate the issue of particle production in
this background. The quantization of the scalar field $\phi$ can be carried out
by choosing enforcing canonical commutation between the fields $\phi_+(\xpo)$
and $\phi_-(\xmo)$ and their canonical conjugate $\p_- \phi_+(\xpo)$ and  $\p_+
\phi_-(\xmo)$. \be
[\phi_+(\xp,\xm,x),\phi^{\dagger}_+(\xp,\xm',x')]=\delta(\xm-\xm')\delta^{(D-2)}(x-x')\ee
\be
[\phi_-(\xp,\xm,x),\phi^{\dagger}_-(\xp',\xm,x')]=\delta(\xp-\xp')\delta^{(D-2)}(x-x')\ee
while $\phi_+,\phi_-$ commute.
This follows by imposing the commutation relations on the past wedge and then
evolving them to the rest of space-time. Following \cite{Tomaras:2001vs} let us
define \be\begin{split}
\phi_+&(\xp,p_+,y)\equiv\int_{\xmo}^{+\infty}d\xm~e^{i\xm
p_+}\phi_+(\xp,\xm,y)\\
=& \int_{\xmo}^{+\infty} dv~e^{ip_+ v }
\label{fourier}
\exp\left(\frac{i}{4p_+} \int_{\xpo}^{\xp} \left(
\frac{m^2}{f^2(u)}-\Delta_T \right)du \right)
\phi_+(\xpo,v,y)\\
+&{ie^{ip_+\xmo}\over p_+}
\times \int_{\xpo}^{\xp} du~\left[\exp\left(\frac{i}{4 p_+}
\int_{u}^{\xp} \left( \frac{m^2}{f^2(u')}-\Delta_T \right) du' \right)\right]
\left( \frac{m^2}{f^2(u)}-\Delta_T \right) \phi_-(u,\xmo,y)\ ,\end{split}
\ee
where as usual $p_+\to p_++i\e$ is understood,
to be the Fourier transform (for $\xmo\to -\infty$) of the field $\phi_+$.
We have also used the relation \be \int_{x}^{\infty}dz~e^{ip~z}={i\over
p}~e^{ip~x} \ee which  follows from (\ref{theta}).
We can now derive the
following relation \be\begin{split}
 -i \p_+ \phi_+(\xp,p_+,y) =& \frac{1}{p_+} \left(
\frac{m^2}{f^2(u)}-\Delta_T \right) \phi_+(\xp,p_+,y) +\\
&+
i{e^{i(p_++i\e)\xmo}\over p_++i\e}\left[{m^2\over f^2(\xp)}-\Delta_T\right]
\phi_-(\xp,\xmo,y)
\end{split}\ee
We are interested in the limit $\epsilon\to 0$, $\xmo\to -\infty$ keeping
$\epsilon \xmo$ constant. Using \be \lim_{x_0^-\to -\infty,\e\to 0^+}
{e^{i(p_++i\e)\xmo}\over p_++i\e} =-2\pi i ~\delta(p_+)\sp \lim_{x_0^-\to
\infty,\e\to 0^+} {e^{i(p_++i\e)\xmo}\over p_++i\e} =0\label{delta}\ee we
obtain
\be\begin{split}
-i \p_+ \phi_+(\xp,p_+,y) = \frac{1}{p_+} & \left(
\frac{m^2}{f^2(u)}-\Delta_T \right) \phi_+(\xp,p_+,y) +\label{ppp}\\
&+2\pi\delta(p_+)~\left[{m^2\over f^2(\xp)}-\Delta_T\right] \phi_-(\xp,\xmo,y)
\end{split}\ee
For $p_+\not =0$ $\phi_+$ is an eigenmode of  the light-cone
Hamiltonian. The sign of its eigenvalue determines whether it is a
creation (resp. annihilation) operator. Since $\Delta_T<0$, the
sign remains that of $p^+$ irrespective of the $f$-profile. We
conclude that there is no particle production of $p_+\not= 0$
states. However, this ceases to be true at $p_+=0$, due to the
second term in (\ref{ppp}). At $p_+=0$ there is no distinction
between particles and anti-particles. Particles can then be
produced, as usual in light-cone quantization.

\section{Holography in the Poincar\'e patch}

In this appendix we provide the calculations for the two- and three-point
amplitudes presented in the main body of the paper. We need the solution to
massive scalar equation with the appropriate (vanishing) boundary condition at
$\xm=-\infty$,
\be \begin{split}
\phi_0(\xp,\xm,x)=&\int dp_+~dp~\chi(p_+,p)\left({f(\xp)\over
f(\xpo)}\right)^{D-2\over 2} \\
&\times \exp\left[i(p_+\xm+p\cdot
x)-i{p^2\over 4p_+}(\xp-\xpo)-i{m^2\over 4p_+}\int_{\xpo}^{\xp}{du\over
f^2(u)}\right]\end{split}
\ee
The renormalized source is $\tilde \chi=\chi~f^{2-D\over
2}(\xpo)$. The two-point amplitude is given by
\bea F_2&=&-{1\over 2}\int d^{D}x
\sqrt{-g} \left( g^{\mu\nu} \p_\mu \phi_0 ~ \p_\nu \phi_0 +m^2
\phi_0^2\right)\nn\\
&=&-{1\over 4}\int {d\xp d\xm d^{D-2}y\over
f^{D-2}(\xp)}\left[4\p_+\phi_0\p_-\phi_0+\p_i\phi_0\p_i\phi_0+{m^2\over
f^2}\phi_0^2\right]=0 \eea Here we have substituted the classical
solution in the action, and found that, after integrating over
$\xm,y$, the resulting integrand is zero. The two-point function
therefore vanishes.

In order
to calculate the three point function, we need the next order solution of the
equation of motion. \be (\Delta-m^2)\phi_1=\phi_0^2 \ee This can be found from
$\phi_0$ using the bulk propagator as \bea \phi_1&=&
\int~f^{-D}(\xp')d\xp'~d\xm'~dx'~G(\xp,\xm,x;{\xp}',{\xm}',x')
\phi_0^2(\xp',\xm',x') \nn\\
&=&\int d\xp'\int dp_+dq_+dpdq~\tilde\chi(p_+,p)\tilde\chi(q_+,q)~H(\xp-\xp'){1
\over 8i(p_++q_+)}\times\\
&&\times f(\xp)^{D-2\over 2} f(\xp')^{(D-6)\over
2}~\exp\left[i(p_++q_+)\xm+i(p+q)\cdot x\right]
\nn\\
&&\times\exp\left[-{i\over 4}\left({p^2\over p_+} (\xp'-\xpo)+{q^2\over
q_+}(\xp'-\xpo)+{(p+q)^2\over p_++q_+}(\xp-\xp')\right)\right]\nn\\
&&
\times\exp\left[-{im^2\over 4}\left({1\over p_+}
\int_{\xpo}^{\xp'}{du\over f^2(u)}+ {1\over q_+}\int_{\xpo}^{\xp'} {du\over
f^2(u)}+{1\over p_++q_+}\int_{\xp'}^{\xp}{du\over f^2(u)}\right)\right]\nn
\eea
The three-point amplitude is given by \be F_3=-\l\int d^{D}x \sqrt{-g} \left(
g^{\mu\nu} \p_\mu \phi_0 ~ \p_\nu \phi_1 +m^2 \phi_0\phi_1 +{1\over
3}\phi_0^3\right)\ee
We substitute again the solutions into the action do the
$\xp,x$ integrals and obtain
\be\begin{split} \int &d^{D}x \sqrt{-g} \left( g^{\mu\nu} \p_\mu
\phi_0 ~ \p_\nu \phi_1 +m^2 \phi_0\phi_1\right)=\\
=&
-{1\over 4}\int dp_+dq_+dr_+dpdqdr~\tilde \chi(p_+,p)\tilde \chi(q_+,q)\tilde
\chi(r_+,r) ~\delta(p_++q_++r_+)~\delta(p+q+r) ~Q_3
\end{split}
\ee
All the terms here cancel point-wise except when a $\p_+$ derivative acts on
$H(\xp-\xp')$ inside $\phi_1$. We also obtain
\be\begin{split} {1\over 3}&\int d^{D}x \sqrt{-g}~\phi_0^3=\\
&= {1\over 6}\int dp_+dq_+dr_+ dp dq dr~\tilde \chi(p_+,p)\tilde
\chi(q_+,q)\tilde \chi(r_+,r) ~\delta(p_++q_++r_+)~\delta(p+q+r)
~Q_3\end{split} \ee with \be
Q_3={(2\pi)^{D-1}}\int_{\xpo}^{x^+_1}du~{f(u)^{D-6\over
2}}\exp\left[-{im^2\over 4}\left({1\over p_+}+{1\over q_+}+{1\over
r_+}\right)\tilde\gamma(u,\xpo)\right]\ee
$$\times
\exp\left[-{i\over 4}\left({p^2\over p_+}+{q^2\over q_+}+{r^2\over
r_+}\right)(u-\xpo)\right]
$$
and \be \tilde\gamma(x^+,\xpo)\equiv \int_{\xpo}^{\xp}{du\over f^2(u)}
\ee
Dropping the external sources and the momentum conserving $\delta$-functions we
obtain for the three-point function \be F_3={\l\over 12} Q_3\ee

\section{Solution of the Ward Identities}

In this section, we study the constraints implied by the algebra of boundary
symmetries $H_D\times H_D$ \bea \label{sboub}
\tilde K&=&\p_- \nn\\
\tilde J_L&=& -\frac{y_1^2+y_2^2}{2}\p_-  + \frac12 (y_1 \p_2-y_2\p_1)+{1\over
8}
(4m^2-\Delta)\p_-^{-1}\nn\\
\tilde J_R&=&-\frac{y_1^2+y_2^2}{2}\p_- - \frac12 (y_1 \p_2-y_2\p_1)+{1\over 8}
(4m^2-\Delta)\p_-^{-1}\nn\\
\tilde P_{iL}&=& y_i \p_-  -\frac{1}{2}\epsilon_{ij} \p_j\\
\tilde P_{iR}&=& y_i \p_-  +\frac{1}{2}\epsilon_{ij} \p_j \nn\eea
on the correlators of the putative dual gauge theory. The Ward
identities for the boundary amplitudes $Q_{n}(p_i,p_+^i)$ take the
form \be \sum_{i=1}^n~({\hat{\cal
L}^{\dagger}_a})_i~Q_{n}(\{p\},\{p_+\})=0 \ee where the operators
are the adjoints of those in (\ref{sboub}). In order for the
symmetry to annihilate the boundary ``vacuum", its generators
should annihilate the part of the classical solution (\ref{phim})
that depends on the vacuum data, $\phi(\xp,\xmo,y)$. It can be
shown that this is true only when $\phi(\xp,\xmo,y)=0$ as we
assumed.

The constraints on the two-point function have already been
discussed in the main text. The $K,P_L^i-P_R^i$ identities for the
three point function imply momentum conservation. Consequently,
\be
F_3(p,q,r;p_+,q_+,r_+)=f_3(p,q,p_+,q_+)\delta(p_++q_++r_+)\delta^{(2)}(p+q+r)
\ee The $P_L^i+P_R^i$, $J_L-J_R$ identities further imply that \be
f_3(p,q,p_+,q_+)=g(p_+,q_+,z)\ee where \be z=\left({p\over
p_+}-{q\over q_+}\right)^2=\left({1\over p_+}+{1\over
q_+}\right)\left({p^2\over p_+}+{q^2\over q_+}+{r^2\over
r_+}\right) \ee Finally, the $J_L+J_R$ identity leads to the
differential equation $g$ \be \left[\p_{z}^2+{1\over
z}\p_z-{B\over 4Az}-{1\over 16A^2}\right]g=0 \ee where
\be
A={1\over p_+}+{1\over q_+}\sp B={m_1^2\over p_+}+{m_2^2\over
q_+}+{m_3^2\over r_+}
\ee
This equation is related to the confluent hypergeometric equation.
The two linearly independent solutions are \be g_+=e^{-{z\over
4A}}F\left({ (1+B\over 2},1,{z\over 2A}\right) \ee
\be g_-=e^{-{z\over 4A}}U\left({ (1+B\over 2},1,{z\over 2A}\right)
\ee where $F(a,b,x)$ is Kummer's confluent hypergeometric
function\be F(a,b,x)\equiv 1+{a\over b}{x\over 1!}+{a(a+1)\over
b(b+1)}{x^2\over 2!}+\cdots \ee and $U(a,b,x)$ is the linearly
independent conjugate Kummer's function. In our case it is
logarithmic \be U(a,1,x)=-{1\over \Gamma(a)}\left[{\p \over \p
a}F(a,1,x)+(2\gamma_E+\psi(a)+\log x)F(a,1,x)\right] \ee We also
have the integral representations
\be F(a,b,x)={\Gamma(b)\over
\Gamma(a)\Gamma(b-a)}\int_{0}^1dt~e^{zt}~t^{a-1}(1-t)^{b-a-1} \ee
\be U(a,b,x)={1\over \Gamma(a)}\int_0^{\infty}dt
~e^{-zt}~t^{a-1}(1+t)^{b-a-1} \ee Using the asymptotic limits \be
F(a,b,x)={\Gamma(b)\over \Gamma(a)}x^{a-b}e^x\left[1+{\cal
O}\left({1\over x}\right)\right] \ee \be
U(a,b,x)=x^{-a}\left[1+{\cal O}\left({1\over x}\right)\right] \ee
we  obtain \be g_{\pm}\sim z^{{-1\pm B\over 2}}~e^{\pm{z\over
4A}}\sp z\to \infty \ee Thus, the general solution to the Ward
identities for the 3-point amplitude is given by \be \label{f3pt}
f_3(p,q;p_+,q_+)=\exp\left[-{z\over
4A}\right]\left[C_+(A,B)F\left({ (1+B\over 2},1,{z\over
2A}\right)+ \right.\ee$$\left. +C_-(A,B)U\left({ (1+B)\over
2},1,{z\over 2A}\right)\right]$$ It is easy to see that our
three-point amplitude for the scalar theory satisfies the
appropriate second order equation and is thus of this form. For
the $N$-point function, the solution to all but the $J_L+J_R$ Ward
identity can be written as \be F_n(p_I,p_+^I)=f_n(z^i,\hat
z^{\alpha},p_+^K)~\delta\left(\sum_{I=1}^n~p_+^I\right)~\delta^{(2)}
\left(\sum_{I=1}^n~p_I\right) \ee where \be z^i={p_1^2\over
p_+^1}+{p_2^2\over p_+^2}+{p_i^2\over p_+^i}\sp i=3,4,\cdots,n \ee
\be \hat z^{\alpha}={p_1^2\over p_+^1}+{p_3^2\over
p_+^3}+{p_{\alpha}^2\over p_+^{\alpha}}\sp \alpha=4,5,\cdots,n \ee
Finally the $J_L+J_R$ Ward identity implies a second order partial
differential equation on the variables $z^i,\hat z^{\alpha}$.

\section{Holography in the harmonic oscillator basis}

In this appendix we translate our holographic derivation of
boundary correlators to the (global) harmonic oscillator basis. At
fixed $p_+$ the transverse coordinates are effectively compact. We
can therefore reduce the wave equation  \`a la Kaluza-Klein. The
analogue of the spherical harmonics are the modes \be
D_{p_+,n_1,n_2}(v,x^i)={1\over \sqrt{2\pi}}
e^{ip_+v}\psi_{n_1}(x^1)\psi_{n_2}(x^2)\ee where $\psi_n$ are the
normalized harmonic oscillator wave-functions with frequency
$\omega =|p_+|/2$, \be \int dvd^2 x~D^*_{p_+,n_1,n_2}(v,x^i)
D^*_{p_+',n_1',n_2'}(v,x^i)=\delta(p_+-p_+')\delta_{n_1,n_1'}
\delta_{n_2,n_2'}\ee We make the decomposition
\be\phi(u,v,x^i)=\sum_{n_{1,2}=0}^{\infty}\int_{-\infty}^{\infty}dp_+~
\xi_{p_+,n_1,n_2}(u)D_{p_+,n_1,n_2}(v,x^i)\label{com}\ .\ee The
Laplacian equation for a complex massive scalar \be
(2\partial_u\partial_v+{|x|^2\over
4}\partial_v^2+\partial_1^2+\partial_2^2-m^2)\phi=0\ee becomes \be
\partial_u \xi_{p_+,n_1,n_2}=-{i\over
2p_+}(m^2+|p_+|(n_1+n_2+1)) \xi_{p_+,n_1,n_2}\ee Consider a
complex scalar scalar field with free action \be S={1\over 2}\int
dudvd^2x[\partial_u\phi\partial_v\phi^*+\partial_u\phi^*\partial_v\phi+{|x|^2\over
4}|\partial_v\phi|^2+|\partial_i\phi|^2+m^2|\phi|^2] \ee We
"compactify" using the decomposition (\ref{com}). For this we need
the matrix elements $\langle m|x^2|n\rangle$ and $\langle
m|\partial_x^2|n\rangle$ for a single harmonic oscillator \bea
\langle m|x^2|n\rangle&=&{1\over 2\omega}X_{m,n}\equiv{1\over
2\omega}
\left[\sqrt{n(n-1)}\delta_{m,n-2}+(2n+1)\delta_{m,n}+\sqrt{(n+1)(n+2)}\delta_{m,n+2}\right]
\nn\\
\langle m|\partial_x^2|n\rangle&=&{\omega\over
2}P_{m,n}\equiv{\omega\over 2}
\left[\sqrt{n(n-1)}\delta_{m,n-2}-(2n+1)\delta_{m,n}+\sqrt{(n+1)(n+2)}\delta_{m,n+2}\right] \nn
\eea 
The action of the symmetry in this basis can be inferred from
the following: For a two-dimensional symmetric harmonic oscillator
we can write the angular momentum as \be L={i\over 2}
(x^1\partial_2-x^2\partial_1)={i\over 2}
(a_2a^{\dagger}_1-a_1a^{\dagger}_2)\ee The spectrum of $L$ for
states $|n,N-n\rangle$ takes the values $-{N\over 2}\leq L\leq
{N\over 2}$ and is integer-spaced. Thus, states can be classified
by their energy $\omega(N+1)$ and the angular momentum $L$. Since
\be [L,a_1\pm ia_2]=\mp{1\over 2}(a_1\pm a_2)\sp
[L,a^{\dagger}_1\pm ia^{\dagger}_2]=\mp{1\over 2}(a^{\dagger}_1\pm
a^{\dagger}_2)\ee we obtain \bea (a_1\pm
ia_2)|N,L\rangle&=&\sqrt{N\pm 2L}|N-1,L\mp{1\over 2}\rangle\\
(a^{\dagger}_1\pm ia^{\dagger}_2)|N,L\rangle&=&\sqrt{N+2\mp
2L}|N+1,L\mp{1\over 2}\rangle\eea where $\langle
N,L|N',L'\rangle=\delta_{N,N'}\delta_{L+L'}$. The action can be
further evaluated using
 \bea \int
dvd^2x~D^*_{p_+,n_1,n_2}\partial_vD_{p_+',n_1',n_2'}&=&ip_+\delta(p_+-p_+')\delta_{n_1,n_1'}
\delta_{n_2,n_2'}\\
\int dvd^2x~{|x|^2\over 4}\partial_v
D^*_{p_+,n_1,n_2}\partial_vD_{p_+',n_1',n_2'}&=&{|p_+|\over 4}
\delta(p_+-p_+')
\left[\delta_{n_1,n_1'}X_{n_2,n_2'}+\delta_{n_2,n_2'}X_{n_1,n_1'}\right]\nn\\
\int dvd^2x~\left[\partial_{x^1}
D^*_{p_+,n_1,n_2}\partial_{x^1}D_{p_+',n_1',n_2'}\right.&+&\left.\partial_{x^2}
D^*_{p_+,n_1,n_2}\partial_{x^2}D_{p_+',n_1',n_2'}\right]\\
&&=-{|p_+|\over
4}\delta(p_+-p_+')\left[\delta_{n_1,n_1'}P_{n_2,n_2'}+
\delta_{n_2,n_2'}P_{n_1,n_1'}\right]\nn\eea Using  the above the
action becomes \be S={1\over
2}\sum_{n_{1,2}=0}^{\infty}\int_{-\infty}^{\infty}dp_+~\int
du~\left[2p_+ Im(\partial_u\xi_{p_+,n_1,n_2}\xi^*_{p_+,n_1,n_2})
+m^2|\xi_{p_+,n_1,n_2}|^2\right]+ \ee
$$+{1\over
8}\sum_{n_{1,2},m_{1,2}=0}^{\infty}\int_{-\infty}^{\infty}dp_+~\int
du~|p_+|\xi^*_{p_+,n_1,n_2}\xi_{p_+,m_1,m_2}\left[\delta_{n_1,m_1}
(X-P)_{n_2,m_2}+\delta_{n_2,m_2}(X-P)_{n_1,m_1}\right]$$
$$={1\over
2}\sum_{n_{1,2}=0}^{\infty}\int_{-\infty}^{\infty}dp_+~\int
du~\left[2p_+ Im(\partial_u\xi_{p_+,n_1,n_2}\xi^*_{p_+,n_1,n_2})
+(m^2+|p_+|(n_1+n_2+1))|\xi_{p_+,n_1,n_2}|^2\right]
$$
Evaluating the action on the classical solution \be
\xi_{p_+,n_1,n_2}(u)=\zeta_{p_+,n_1,n_2}\exp\left[-{i\over
2p_+}(m^2+|p_+|(n_1+n_2+1))u\right]\ee we obtain \be S={V_u\over
2}\sum_{n_{1,2}=0}^{\infty}\int_{-\infty}^{\infty}dp_+
\left[(2p_+p_-+m^2+2|p_+|(n_1+n_2+1))|\zeta_{p_+,n_1,n_2}|^2\right]=0\ee
where $V_u$ is the (infinite) volume of the $u$-line and \be
p_-=-{1\over 2p_+}(m^2+|p_+|(n_1+n_2+1))\ee Thus, as in Appendix
C, two-point interactions of the boundary sources vanish.

We will now consider the cubic interaction of the form \be
S_{int}=g\int dudvd^2x~\phi_1\phi_2\phi_3 \ee where $\phi_i$ are
real scalars with masses $m_i$. Define the overlap integral of
harmonic oscillator wave-functions
\be\int_{-\infty}^{\infty}dx~\psi^{\omega_1}_{n_1}(x)~\psi^{\omega_2}_{n_2}
(x)\psi^{\omega_3}_{n_3}(x)=C_{n_1,n_2,n_3}(\omega_1,\omega_2,\omega_3)
\ee
It can be evaluated from the generating function
\bea
C(t_1,t_2,t_3;\omega_1,\omega_2,\omega_3)&\equiv&
\sum_{n_1,n_2,n_3=0}^{\infty}{(\sqrt{2}t_1)^{n_1}(\sqrt{2}t_2)^{n_2}(\sqrt{2}t_3)^{n_3}\over
\sqrt{n_1!~n_2!~n_3!}}C_{n_1,n_2,n_3}(\omega_1,\omega_2,\omega_3)\nn\\
=\left({4\omega_1\omega_2\omega_3\over
\pi(\omega_1+\omega_2+\omega_3)^2}\right)^{1\over
4}&&\exp\left[2{(\sqrt{\omega_1}t_1+\sqrt{\omega_2}t_2+\sqrt{\omega_3}t_3)^2\over
\omega_1+\omega_2+\omega_3}-{1\over
2}(t_1^2+t_2^2+t_3^2)\nn\right]
\eea
We obtain for the three-point amplitude
 \bea
 S_{int}&=&g\sqrt{2\pi}\sum_{n^i_{1,2}=0}^{\infty}\int~
 \left[\prod_{i=1}^3~dp_+^i~\zeta_{p_+^i,n_1^i,n^i_2}\right]
\delta (p_-^1+p_-^2+p_-^3) \delta (p_+^1+p_+^2+p_+^3)\times\nn\\
&\times &
C_{n^1_1,n_1^2,n_1^3}(|p_+^1|,|p_+^2|,|p_+^3|)C_{n^1_2,n_2^2,n_2^3}(|p_+^1|,|p_+^2|,|p_+^3|)
\eea
Higher point amplitudes can be computed along similar lines.

\end{document}